\documentclass[twocolumn,aps,prb,showpacs,showkeys,superscriptaddress,floatfix]{revtex4-1}
\usepackage[T1]{fontenc}
\usepackage[utf8]{inputenc}
\usepackage{amsmath}
\usepackage{amssymb}
\usepackage{graphicx}
\usepackage{esint}
\usepackage{color}

\begin{document}

\title{Interface excitons at lateral heterojunctions in monolayer semiconductors}

\author{Ka Wai Lau}

\affiliation{Department of Physics, and Center for Theoretical and Computational
Physics, The University of Hong Kong, China}

\author{Calvin}

\affiliation{Department of Physics, and Center for Theoretical and Computational
Physics, The University of Hong Kong, China}

\author{Zhirui Gong}

\email{gongzr@szu.edu}

\affiliation{College of Physics and Energy, Shenzhen University, Shenzhen 518060,
P. R. China}

\affiliation{Department of Physics, and Center for Theoretical and Computational
Physics, The University of Hong Kong, China}

\author{Hongyi Yu}

\affiliation{Department of Physics, and Center for Theoretical and Computational
Physics, The University of Hong Kong, China}

\author{Wang Yao}

\affiliation{Department of Physics, and Center for Theoretical and Computational
Physics, The University of Hong Kong, China}
\begin{abstract}
We study the interface exciton at lateral type II heterojunctions of
monolayer transition metal dichalcogenides (TMDs), where the electron and hole prefer to stay at complementary
sides of the junction. We find that the 1D interface exciton
has giant binding energy in the same order as 2D excitons in pristine
monolayer TMDs although the effective radius (electron-hole seperation) of interface
exciton is much larger than that of 2D excitons. The binding energy, exciton radius and optical dipole
strongly depends on the band offset at the junction.
The inter-valley coupling induced by the electron-hole Coulomb exchange interaction and the quantum confinement effect at interface of a closed triangular shape are also investigated. Small triangles realize 0D quantum dot confinement of excitons, and we find a transition from non-degenerate ground state to degenerate ones when the size of the triangle varies. Our findings may facilitate
the implementation of the optoelectronic devices based on the lateral
heterojunction structures in monolayer semiconductors.
\end{abstract}

\pacs{71.35.-y,73.22.-f,31.15.ve}

\date{\today}

\maketitle

\section{Introduction}

Heterostructures between conventional three-dimensional (3D) semiconductors
has inspired the inventions of the modern electronic devices such
as high speed transistors\cite{Takashi80}, diode lasers\cite{Zory93},
light-emitting diodes\cite{Morkoc95} and solar cells\cite{Yablonovitch86}.
Thanks to the development of nanotechnology, we are able to engineer heterostructures on the nanoscale for high-speed opto-electronic devices. In III-V and II-VI
semiconductors, various nanoscale heterostructures such as quantum
wells, superlattices, and core-shell nanodots and nanowire have been
widely studied\cite{Bastard88,Grundman98,Li11}. Emerged as a new
class of semiconductors in the two-dimensional (2D) limit\cite{Liu15,Mak10,Splendiani10,Wang12,Xu14},
monolayers of group-VIB transition metal dichalcogenides (TMDs) possess
visible range direct gap, excotic properties associated with valley
degeneracy, and new geometries for realizing various heterostructures,
which provide new platforms to study the physics and applications
at semiconductor heterostructures\cite{Cao12,Jones13,Mak12,Yao08}.
By stacking different TMDs monolayers which are then bound together
by the weak interlayer Van der Waals forces, vertical heterostructures
have been realized recently, e.g. $\mathrm{MoX_{2}}$/$\mathrm{WX_{2}}$
(X=Se, S) heterobilayers\cite{Cheng14,Chiu14,Fang14,Furchi14,Hong14,Lee14,Rivera15,Rivera16}
which can be analogs of the III-V semiconductor double heterojunctionss.

Besides the vertical heterostructures, two-dimensional materials also make possible heterostructures of a unique planar geometry. Two different TMDs seamlessly connected
in a single monolayer has been realized experimentally already\cite{Chen17,Bogaert16,Chen15a,Chen15b,Duan14,Gong15,Gong14,Huang14,Li15,Samani15,Tizei15,Zhang15}.
A more recently development is on the growth of various lateral
heterostructures, multi-heterostructures and superlattices for TMDs~\cite{Zhang17}.
The possibility to form atomically sharp and straight lateral interface of different compounds~\cite{Li15,Zhang17} point to exciting opportunities towards device applications based on the lateral heterojunctions, as well as a new geometry to realize quantum wires and even quantum dots in the monolayer semiconductors. The lateral
heterojunctions can also be realized in an alternative way, by electrostatic gating to define lateral p-n junctions ~\cite{Ross14,Baugher14,Pospischil14}. The recent development shows that the width of the electric gate in
the monolayer MoS2 can be narrowed down to 1nm by using a single-walled
carbon nanotube as the gate electrode~\cite{Desai16}.

In most of the vertical and lateral heterostructures formed between
different TMDs monolayers, they feature a type-II band alignment,
where the conduction and valence band edge locate in different TMDs.
The strong Coulomb interaction binds electron and hole to form exciton at the interface.  In contrast to 2D exciton formed in
pristine monolayer TMD, the electron and hole at the interface
will be spatially separated because of the type-II band alignment,
and such an interface exciton can have lower energy, being an excitonic
ground state in the heterostructures. The properties of such interface
excitons can be essential to determine the optical response of the
lateral heterostructures of TMDs. In vertical heterojunctions $\mathrm{MoX_{2}}$/$\mathrm{WX_{2}}$
heterobilayers, such interface exciton has already been investigated
theoretically and experimentally\cite{Rivera15,Rivera16,Zande14,Yu15,Yu17,Wu18,Yu18}.
Due to the spatial separation of electron and hole, interlayer excitons in $\mathrm{MoX_{2}}$/$\mathrm{WX_{2}}$
heterobilayers have shown long lifetime exceeding nanoseconds \cite{Rivera15,Rivera16}
and electro-statically tunable resonance~\cite{Jones14} which are
highly desirable for the realization of excitonic circuits and condensation~\cite{Eisenstein04,High08}.
And interestingly, the inevitable twisting and lattice
mismatch in the heterobilayers can give rise to novel light coupling
properties~\cite{Zande14,Yu15,Yu17,Wu18,Yu18}. Albeit the novel and appealing properties
discovered, the interface excitons in the heterobilayers of 2D semiconductors
are analogues of those in the conventional heterostructures bulk semiconductors,
for example the spatially indirect excitons in III-V double quantum
well. The realization of lateral heterostructures in monolayer TMDs opens up new opportunity to extend the study of interface exciton from two-dimensional interface to the one-dimensional (1D) interface.
The 1D interface exciton mode may shed light on novel optoelectronic
devices based on these atomically thin 2D lateral heterostructures.
Moreover, such 1D interface excitons may also become relevant in lateral p-n junctions in monolayers TMDs\cite{Baugher14,Pospischil14,Ross14}.

Here, we theoretically study the interface exciton states at lateral
heterojunctions of the monolayer TMDs. The physical properties of one-dimensional type-II interface exciton such
as the binding energy, exciton radius (i.e. electron-hole separation), longitudinal-transverse splitting by the electron-hole exchange and optical dipole are investigated as a function of band offset at the interface. We adopted two different approaches to
calculate the interface exciton states. One approach bases on a real-space
tight binding (TB) model, and the other approach uses the perturbation
expansion in a hydrogen-like basis ineffective mass approximation. The
numerical study shows with the increase of the band offset at the interface, the exciton radius grows and can become several times larger than that of the 2D excitons in homogeneous monolayer TMDs.  In the meantime, the decrease in the exciton binding energy is not as significant, remaining in the same order as the 2D exciton, because of  the weaker
screening of Coulomb interaction as electron-hole separation increases.
Due to spatial indirect nature of interface exciton, the optical transition
dipole decreases fast with the increase of band offset, which, at a typical band offset of ~ 300 meV, is about one order of magnitude smaller than that of 2D exciton . We also investigated lateral heterostructures with a closed triangular shaped interface which effectively realize a 0D quantum dot confinement of exciton. Such quantum dot uniquely features the quantum confinement of one carrier by the band offset of the interface, and binding of the other carrier in the proximity exterior by the strong Coulomb.  We find two distinct scenarios of energy level schemes and valley optical selection rules of the interface exciton at small and large quantum dot size respectively, which can be exploited for optical quantum controls.

The paper is organized as follows. In section II, we introduce the
Hamiltonian of the exciton of lateral structures in the effective
mass approximation. We study the interface exciton at the 1D p-n and p-n-p heterojunctions of monolayer semiconductors in Sec. III. The numerical results of the physical observables of interface exciton are also shown in section III. In Sec. IV, we show the numerical calculation of the interface exciton at the 0D quantum dot type triangular lateral heterostructure. We conclude in section V.

\section{Hamiltonian in the effective mass approximation}

The three-band model involving all d orbitals of the transition metal
atom is usually applied to describe the single electron in the monolayer
TMDs associated with valley index~\cite{Liu15}. In the low energy
excitation limit where only the electron in the vicinity of the valance
band edge is excited by light field to the vicinity of the conduction
band edge, both the electron in the conduction band and the hole left
in the valence band can be approximately described by the effective
mass model. In this sense, the periodic parts of the electron and
hole Bloch wavefunctions are omitted and only the profiles of the
electron and hole Bloch wavefunctions are taken into consideration
in the following discussion. Together with the attractive Coulomb
interaction and the lattice potentials, the type II interface exciton at the
interface can be described by the following Hamiltonian
\begin{equation}
H=-\frac{\hbar^{2}}{2m_{e}}\nabla_{\mathbf{r}_{e}}^{2}-\frac{\hbar^{2}}{2m_{h}}\nabla_{\mathbf{r_{\mathrm{h}}}}^{2}+V_{\mathrm{C}}\left(\left|\mathbf{r}_{e}-\mathbf{r}_{h}\right|\right)+V_{\mathrm{e}}(\mathbf{r}_{e})+V_{\mathrm{h}}(\mathbf{r}_{h}),\label{eq:two-body}
\end{equation}
where $m_{e}$ $\left(m_{h}\right)$ is the electron (hole) effective
mass, and $\mathbf{r}_{e}$ ($\mathbf{r}_{h}$) denotes the position
coordinates of the electron (hole). The lattice potentials of electron
and hole $V_{\mathrm{e}}(\mathbf{r}_{e})$ and $V_{\mathrm{h}}(\mathbf{r}_{h})$
depends on the different geometries of the lateral heterostructures.

Here, the Coulomb interaction $V_{\mathrm{C}}\left(\left|\mathbf{r}_{e}-\mathbf{r}_{h}\right|\right)$
between the electron and hole in the 2D limit reads~\cite{Keldysh78,Cudazzo11}
\begin{eqnarray}
V_{\mathrm{C}}(r) & = & -\frac{e^{2}\pi}{2r_{0}}\left(H_{0}(\frac{r}{r_{0}})-Y_{0}(\frac{r}{r_{0}})\right),\label{eq:Veff}
\end{eqnarray}
where $H_{n}$ and $Y_{n}$ denote Struve Function and Bessel Function
of the Second Kind respectively. The former researches demonstrated
that in monolayer TMDs the quasi-2D geometry leads to a distance-dependent
effective dielectric screening \cite{Berkelbach13,Chernikov14,He14,Qiu13}.
For monolayer TMDs, the parameter $r_{0}$ is in the order of a few
nm, which is comparable to the Bohr radius of a free 2D exciton \cite{Berkelbach13,Chernikov14}.

\section{Interface exciton at 1D p-n and p-n-p heterojunctions}

\begin{figure}[ptb]
\includegraphics[width=3.5in]{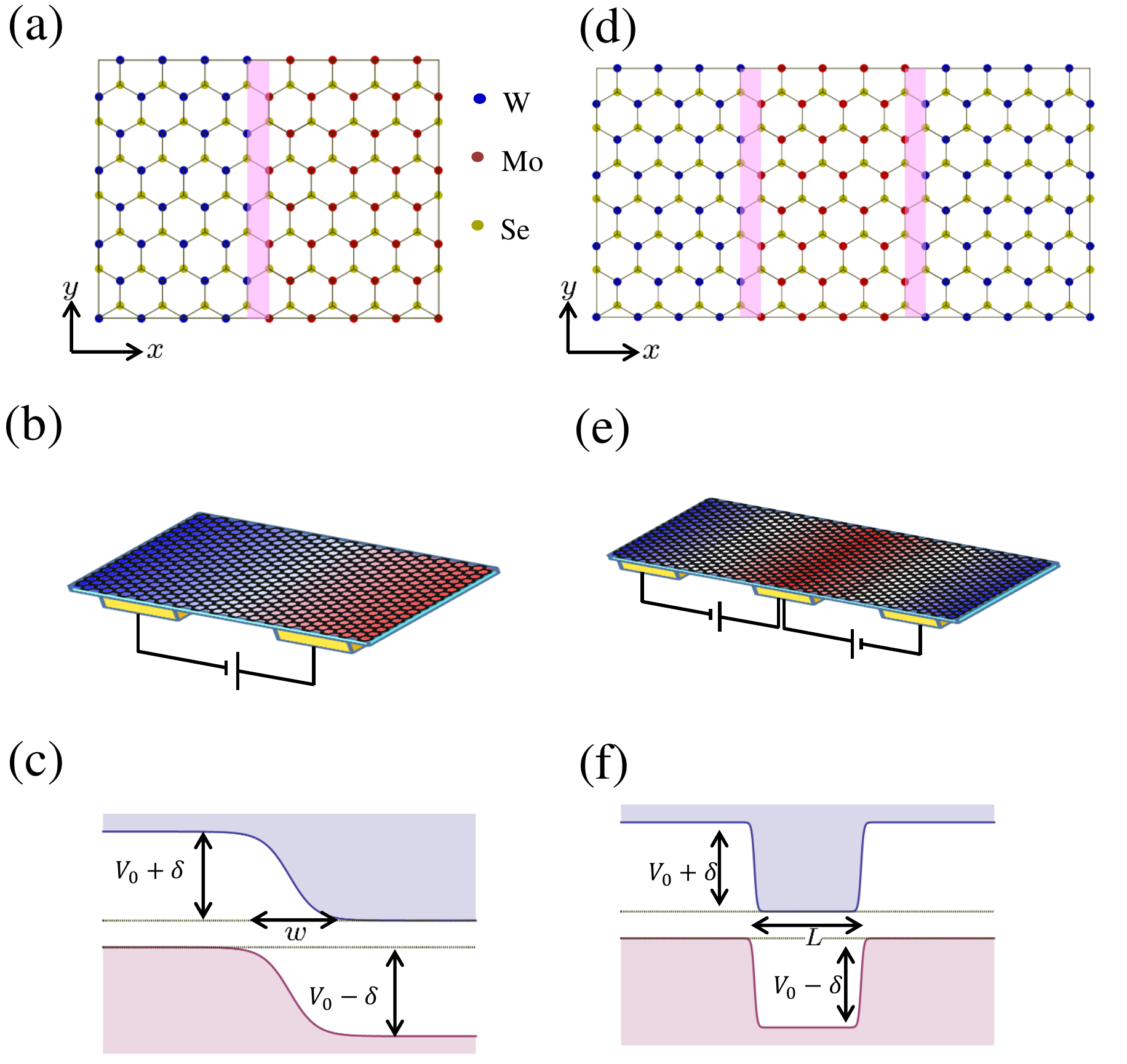}

\caption{(Color online) (a) (d) Schematics of the single- and double- heterojunctions formed between $\mathrm{MoSe}_{2}$ and $\mathrm{WSe}_{2}$.
The blue, red and yellow spheres respectively denote $\mathrm{W}$,
$\mathrm{Mo}$ and $\mathrm{Se}$ atoms. The purple shadow regions denotes the type II interfaces. (b) (e) Schematics of the monolayer TMDs p-n junction defined by the gate voltage. The red and blue regions denote the n and p regions,
respectively. (c) (f) Illustration of the band edge profile across the single lateral junction in (a) or (b), and the double lateral junctions in (d) or (e). Here, $w$ is the width of the interface, which is zero for the case of (a) and (d), and has a finite value for the gate defined junctions in (b) and (e). $V_{0}$ is the average
lattice potential of electron and hole in (a) or (d) and gate voltage in (b) or (e). Here, $\delta$ characterizes the difference of the band offsets for electron and hole.}

\label{fig:fig1}
\end{figure}


\subsection{Type II interface in monolayer TMDs}

A lateral type-II interface in monolayer TMDs can be implemented in
two setups. The first is a lateral heterojunction seamlessly formed
between different TMDs \cite{Duan14,Gong15,Gong14,Huang14}
as shown in Fig. \ref{fig:fig1}(a) and Fig. \ref{fig:fig1}(d), In
such cases, the type-II interface is atomically sharp. The conduction
and valence band edges as functions of position are regarded as the
step functions. The other setup shown in Fig. \ref{fig:fig1}(b) and
Fig. \ref{fig:fig1}(e) is a lateral p-n or p-n-p junctions electrostatically
created in a monolayer TMD by separate back gates, which has been
studied experimentally \cite{Baugher14,Pospischil14,Ross14}. Such setup realizes a gentle type-II interface with a finite width of the interfaces $w$.

We are interested in the binding energy and wavefunction of the interface exciton ground state, which determines the stability and
optoelectronics properties of the interface exciton. By the interface
potentials, electron and hole prefer to stay at complementary sides of
the interface, while the Coulomb interaction $V_{\mathrm{C}}\left(\left|\mathbf{r}_{e}-\mathbf{r}_{h}\right|\right)$
attempts to bind the electron and hole. The properties of the interface
exciton therefore depends on the competition of the band offset
and the Coulomb interaction, which are then tunable by the width $w$
of the interface and the magnitude of the conduction and valence band
edge offsets  $V_{0}$ . The effective dielectric screening of Coulomb interaction
varies with distance, and for large distance between electron and
hole the screening effect is substantially reduced. As we will show,
this is important for the interface exciton to have strong binding energy,
even though the spatial separation between theelectron and hole is much larger than the Bohr radius of the 2D exciton.

\subsection{Solving the eigen problem using Bohr-Oppeheimer Approximation}

Since $V_{\mathrm{I}}(\mathbf{r}_{e},\mathbf{r}_{h})=V_{\mathrm{e}}(\mathbf{r}_{e})+V_{\mathrm{h}}(\mathbf{r}_{h})$
possesses translational symmetry while Coulomb interaction possesses
rotational symmetry, incompatible symmetries make it impossible to obtain analytical solutions for the Schrödinger equation governed by the Hamiltonian in Eq.(\ref{eq:two-body}).
We rewrite the above Hamiltonian with the center-of-mass motion and
relative motion of the electron-hole pair as
\begin{eqnarray}
H & = & -\frac{\hbar^{2}}{2M}\nabla_{\mathbf{R}}^{2}-\frac{\hbar^{2}}{2\mu}\nabla_{\mathbf{r}}^{2}+V_{\mathrm{C}}\left(r\right)+V_{\mathrm{I}}\left(\mathbf{R},\mathbf{r}\right),\label{eq:rewritten Hamil-1}
\end{eqnarray}
where the center-of-mass and relative space coordinates are
\begin{equation}
\begin{cases}
\mathbf{R}= & \frac{1}{M}\left(m_{e}\mathbf{r}_{e}+m_{h}\mathbf{r}_{h}\right)\\
\mathbf{r}= & \mathbf{r}_{e}-\mathbf{r}_{h}
\end{cases}
\end{equation}
with total mass $M=m_{e}+m_{h}$ and reduced mass $\mu=m_{e}m_{h}/\left(m_{e}+m_{h}\right)$.
Due to the 2D nature of TMDs, these coordinates only have two components which means $\mathbf{R}=(X,Y)$ and $\mathbf{r}=(x,y)$. Obviously, the total mass is at least four times greater than the reduced mass ($M\geq 4\mu$), which implies that the center-of-mass motion is a relatively slow one in comparison with the relative motion. Under this circumstance, we can apply the Bohr-Oppenheimer Approximation (BOA) here and in zeroth order BOA the eigen-wavefunction is a product state as
\begin{equation}
\Phi\left(\mathbf{R},\mathbf{r}\right)=\Psi\left(\mathbf{R}\right)\Theta\left(\mathbf{R},\mathbf{r}\right).\label{eq:naive}
\end{equation}
For lateral heterojunctions $V_{\mathrm{I}}(\mathbf{r}_{e},\mathbf{r}_{h})$
possesses translational symmetry along $y$-direction as shown in
Fig. \ref{fig:fig1}(a) and Fig. \ref{fig:fig1}(d). The interface
potential is numerically modelled with $V_{\mathrm{e}}(x_{e})=\frac{V_{0}+\delta}{2}(1-tanh\left(\frac{x_{e}}{w}\right))$
and $V_{\mathrm{h}}(x_{h})=-\frac{V_{0}-\delta}{2}(1-tanh\left(\frac{x_{h}}{w}\right))$,
where $w$ is the width of the interface which characterizing the sharpness of the band offset. As
$V_{\mathrm{I}}(\mathbf{R},\mathbf{r})$ is independent of
$Y$, the envelope function remains to be a plane wave in $Y$-direction,
so we rewrite the center-of-mass motion part as $\Psi\left(\mathbf{R}\right)=\Psi\left(X\right)e^{iP_{Y}Y}$.
Since $P_{Y}$ stands for the $y$-component wave vector corresponding
to a kinetic energy $\hbar^{2}P_{Y}^{2}/2M$, obviously $P_{Y}=0$
for the ground state of type-II interface exciton. Then the corresponding
Schrödinger's equations for the relative motion and center-of-mass
motion read

\begin{align}
\left[(-\frac{\hbar^{2}}{2\mu}\nabla_{\mathbf{r}}^{2}+V_{\mathrm{C}}(r)+V_{\mathrm{I}}\left(X,\mathbf{r}\right)\right]\Theta\left(X,\mathbf{r}\right) & =E\left(X\right)\Theta\left(X,\mathbf{r}\right),\label{eq:4}\\
\left[-\frac{\hbar^{2}}{2M}\frac{\partial^{2}}{\partial X^{2}}+E\left(X\right)\right]\Psi\left(X\right) & =E_{g}\Psi\left(X\right).\label{eq:5}
\end{align}
The energy $E\left(X\right)$ plays the role of an effective potential
in Eq. (\ref{eq:5}) which leads to the ground state $\Psi\left(X\right)\Theta\left(X,\mathbf{r}\right)$
of type-II interface exciton with corresponding ground state energy
$E_{g}$. To numerically solve Eq.(\ref{eq:4}), we adopted two different
approaches: one is the solution based on a real-space tight binding
model for the relative part Hamiltonian $H_{r}=-\frac{\hbar^{2}}{2\mu}\nabla_{\mathbf{r}}^{2}+V_{\mathrm{C}}(r)+V_{\mathrm{I}}\left(X,\mathbf{r}\right)$,
and the other is a perturbative expansion of $H_{r}$ with a hydrogen-like basis of the effective mass model. Details of both approaches can be found in the numerical results section.

\subsection{Physical observables and the electron-hole overlap}

Before we present the numerical results we would like to introduce
several important physical observables first. When applying BOA and
obtaining the eigen-wavefunction of type-II interface excitons, we
can straightforwardly calculate the binding energy, effective radius, optical dipole and the intervalley coupling of the interface excitons\cite{Yu14}.

The binding energy of type-II interface exciton is defined as $E_{b}=E_{f}-E_{g}$,
where $E_{f}$ is the energy of a non-interacting electron-hole pair
at the interface. The effective radius is straightforwardly calculated as
\begin{equation}
a_{b}=\sqrt{\iint d\mathbf{r}dX\mathbf{r}^{2}\left|\Psi\left(X\right)\Theta\left(X,\mathbf{r}\right)\right|^{2}},
\end{equation}
which measures the spatial separation between the electron and hole.

Another important observable is the optical dipole defined as
\begin{equation}
D=\left|D_{0}\int dX\Psi\left(X\right)\Theta\left(X,\mathbf{0}\right)\right|,\label{eq:OD}
\end{equation}
which relates to the lifetime of type-II interface exciton in TMDs.
Here, $D_{0}$ is the interband transition dipole element between
the conduction band and valence band for 2D exciton~\cite{Citrin93}. In contrast with
the optical dipole of 2D excitons, there is an additional integral
over $X$ direction whose value depends on the wavefunction profile
along $X$ direction. Additionally, $\mathbf{r}=0$ in the relative
part of the wavefunction $\Theta\left(X,\mathbf{0}\right)$ indicates
that the recombination of the electron and the hole in the exciton occurs
only when they exactly locates at the same position. In this sense, the
electron and hole will become harder to recombine with each other and
thus results in a longer lifetime due to the decreased optical dipole.
As we will show that below the amount of overlap between the electron
and hole can be controlled by $V_{\mathrm{I}}(\mathbf{R},\mathbf{r})$.

As we can see, the electron-hole overlap plays an important role in
Eq.(\ref{eq:OD}). For a ground state of the 2D exciton, it closely resembles
s-orbitals, so a very large optical dipole $D_{0}$ is expected. However
for large $V_{0}$ the wavefunction
overlap between the electron and hole is greatly reduced. Besides the separation of the electron-hole pair and optical
dipole $D$, the electron-hole overlap also affects other properties
of the type-II interface exciton such as intervalley coupling induced
by Coulomb exchange interaction\cite{Yu14}.

It was proved that under broken 3-fold rotation symmetry in a monolayer
TMDs, the excitonic spectrum could have a finite valley exchange interaction
even in the ground state which is induced by exchange Coulomb interaction
between electrons and holes\cite{Yu14}. In the presence of $V_{\mathrm{I}}(\mathbf{R},\mathbf{r})$,
translational symmetry is only preserved in the $y$-direction, breaking
the 3-fold rotation symmetry. Thus lateral heterojunctions not only
decrease the electron-hole overlap, but also results in a non-vanishing
valley-exchange term $J$. Such term opens a coupling channel between $+K$
and $-K$ valleys which is normally suppressed in monolayer TMDs due
to large momentum difference. In a quasi-1D system, the intervalley
coupling strength is written as

\begin{equation}
J=(\frac{at}{E_{g}})^{2}\sum_{P_{X}}V_{\mathrm{C}}(P_{X},P_{Y}=0)P_{X}^{2}\left|\psi(P_{X})\right|^{2},\label{eq:JLR}
\end{equation}
where $a$ is the lattice constant, $t$ is the hopping constant,
and $V_{\mathrm{C}}(P_{X},P_{Y})$ is Coulomb interaction in the momentum
space. Here, $\psi(P_{X})=\frac{1}{\sqrt{L_{X}}}\sum_{X}\Phi\left(X,\mathbf{r}=0\right)exp(iP_{X}X)$
is the electron-hole overlap in the $X-$ component momentum space.
Since the electron-hole overlap is controlled by the strength of the
band offset, a tunable intervalley coupling is expected in
the lateral heterojunction in TMDs.

It is important to note that the above described zeroth-order Bohr-Oppenheimer
Approximation is valid only in the adiabatic limit where the gradient
of the band offset caused transition probability is much smaller
than the energy level spacing between the ground state and any excited
state in Eq. (\ref{eq:4}). Detailed justification shall be referred
to the appendix or literature about generalized Bohr-Oppenheimer Approximation
(BOA) \cite{Kendricka02,Sun90}. For the eigen-problem of type-II
interface exciton in TMDs, we will numerically justify that the
zeroth order BOA is sufficient.

We will take $\mathrm{MoSe_{2}}$/$\mathrm{WSe_{2}}$ heterojunctions
as our example in subsequent sections of type-II interface excitons. It is
trivial to generalize our method to other sharp TMDs lateral interfaces.

\subsection{Numerical Results Based on TB Model}

In order to obtain the TB model, we discretize Eq.(\ref{eq:4}) in
the real space. We take unit in $x$-direction as $a$ and
$y$-direction as $\frac{\sqrt{3}}{2}a$, where $a=3.325\textrm{\AA}$
is the lattice constant. The lattice constant of $\mathrm{WSe_{2}}$
and $\mathrm{MoSe_{2}}$ closely matches so it is legitimate to assume
the same lattice constant across the heterojunctions \cite{Furchi14,Xiao12}.
A 72 x 84 supercell and the open boundary conditions for both directions
are taken into consideration. We consider an armchair interface in
the following calculation, while it will give almost the identical
results when changing the armchair edge to a zigzag one. Previous
studies \cite{Xiao12,Liu13} show that the conduction and valence
bands are accurately described by $d$-orbitals of the metal atoms,
while the orbits of the chalcogenides play a minor role. Hence we
only consider the metal atoms in our TB model, and the nearest-neighbour
hopping $t=-\hbar^{2}/3a^{2}\mu$ between metal atoms. The one of
the advantages of applying TB model is that $V_{\mathrm{C}}\left(r\right)$
and $V_{\mathrm{I}}\left(\mathbf{R},\mathbf{r}\right)$ are exactly
diagonalized. It is noticed that the on-site electrostatic energies
$U=V_{C}(r=0)$ is divergent. Since for a type II alignment the electron
and hole barely can occupy the same site, a large value of the on-site
electrostatic energies U is assumed in our calculation in order to
make the calculation convergent.

The effective masses of the electron and hole are chosen as $m_{e}=m_{h}=0.32m_{0}$
and thus the reduced mass is $\mu=0.16m_{0}$ with free electron mass
$m_{0}$. The width of the interface is chosen as $w=0.1a$ to model
a very sharp band offset in order to simulate the single- or double-heterojunctions. Here, $r_{0}$ in the effective Coulomb interaction
is chosen as $r_{0}=75$ $\textrm{\AA}$.\cite{Chernikov14} A symmetric
heterojunction ($\delta=0$) is considered unless otherwise specified.

\begin{figure}[tp]
\includegraphics[width=3.5in]{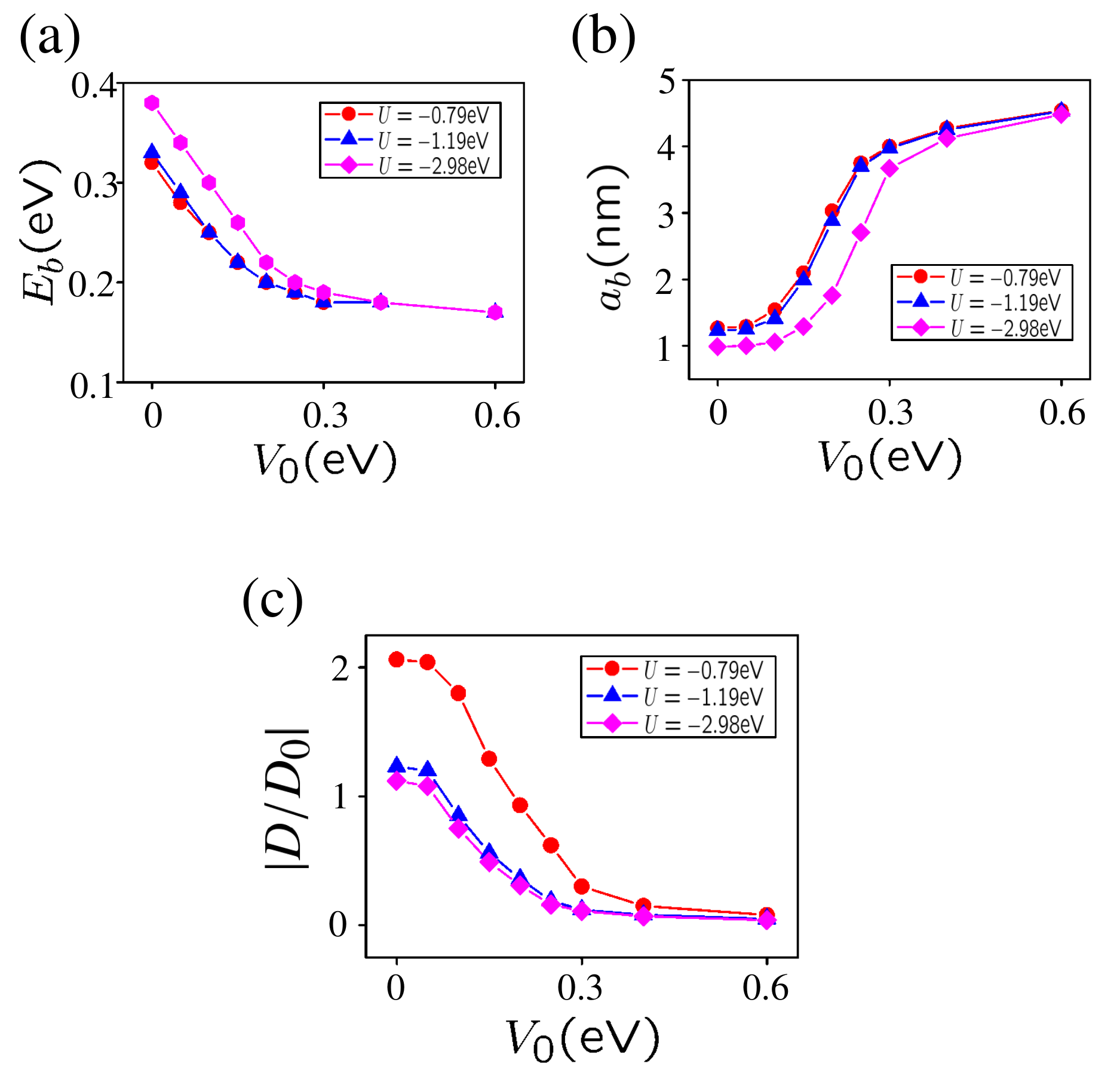}

\caption{(a) The binding energy $E_{b}$,
(b) the exciton radius $a_{b}$, and (c) optical dipole $|D/D_{0}|$ versus
the strength of the band offset $V_{0}$ for different on-site
Coulomb potential $U$. The red sphere, blue triangle and magenta
diamond symbols respectively represent results for different on-site
electrostatic energy $U=-0.79,-1.19$ and $-2.98$eV. The physical
observables converge to the same value at high voltage $V_{0}$ regardless
of $U$, showing the spatial separation nature of the interface exciton for large $V_{0}$. There are two characteristic behaviours, in the regime
of small band offset ($V_{0}\ll0.1$ eV) and large interface
potential ($V_{0}>0.1$ eV), which reflects the competition between
Coulomb interaction and the band offset. See text for details. }

\label{fig:fig2}
\end{figure}

By solving Eg.(\ref{eq:5}), the binding
energy $E_{b}$, effective radius $a_{b}$ and optical dipole $D$ versus different strength of the band offset $V_{0}$ are shown
in Fig. \ref{fig:fig2}. We also depict those physical observables
when different on-site electrostatic energies $U$ are chosen. The
red sphere, blue triangle and magenta diamond symbols respectively
represent $U=-0.79,-1.19$ and $-2.98$ eV. The physical observables
converge to the same value at high voltage $V_{0}$ regardless of
$U$, which actually implies that at large $V_{0}$ the electron and
hole are well separated and thus there is almost no on-site electrostatic
energy contribution in $E_{b}$. Basically there are two characteristic
behaviours, the regime of small band offset ($V_{0}<0.1$
eV) and large band offset ($V_{0}>0.4$ eV). This reflects
the competition between Coulomb interaction and the band offset.
For small band offset $V_{0}$, $V_{\mathrm{C}}\left(r\right)$
dominates over $V_{\mathrm{I}}\left(\mathbf{R},\mathbf{r}\right)$
so the exciton ground state is almost equivalent to a 2D exciton while
$V_{\mathrm{I}}\left(\mathbf{R},\mathbf{r}\right)$ is regarded as
a perturbation term. Therefore, the physical observables of type-II
interface exciton are almost the same as the the ones for 2D exciton
in this regime. However, for sufficiently large $V_{0}$, $V_{\mathrm{I}}\left(\mathbf{R},\mathbf{r}\right)$
dominates over $V_{\mathrm{C}}\left(r\right)$. The effective radius $a_{b}$ shows a rapid rise while $D$ drops dramatically as the band offset increases.
In this sense, we can control physical properties of type-II interface
exciton by adjusting the band offset.


\begin{figure}
\includegraphics[width=3.5in]{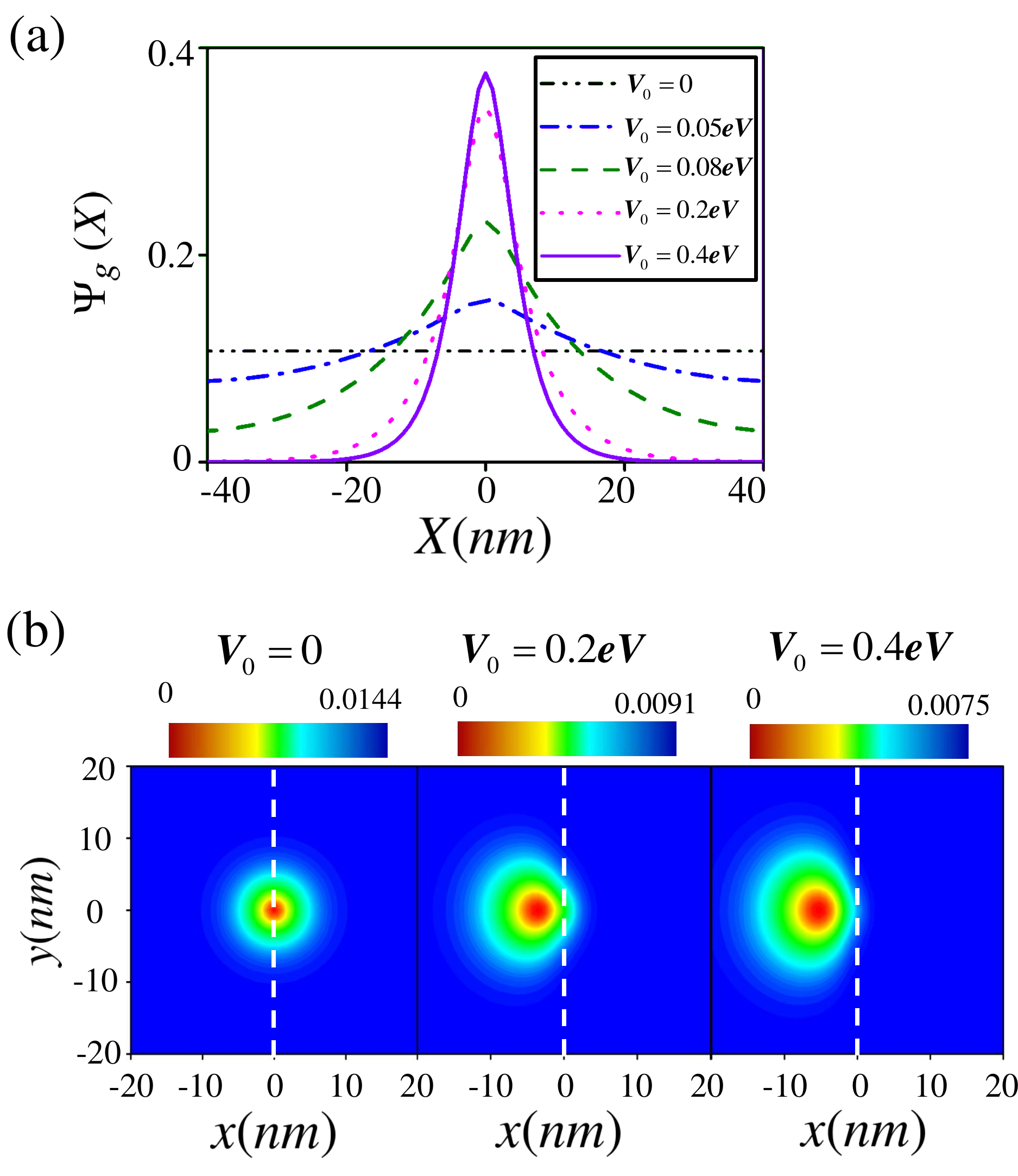}

\caption{(Color online) (a) Center-of-mass wave function, and (b) relative
motion spatial probability distribution at different $V_{0}$. The white dashed line denotes the central position of
the interface. From (a) we see that the wavefunction undergoes a transition
from an extended state to a localized state as $V_{0}$ increases,
demonstrating the competition between the Coulomb interaction and the band offset.
From (b), we see that for larger $V_{0}$ the electron-hole pair tends
to be farer apart and electron-hole overlap is greatly reduced.}

\label{fig:fig3}
\end{figure}



\begin{figure}[ptb]
\includegraphics[width=3.4in]{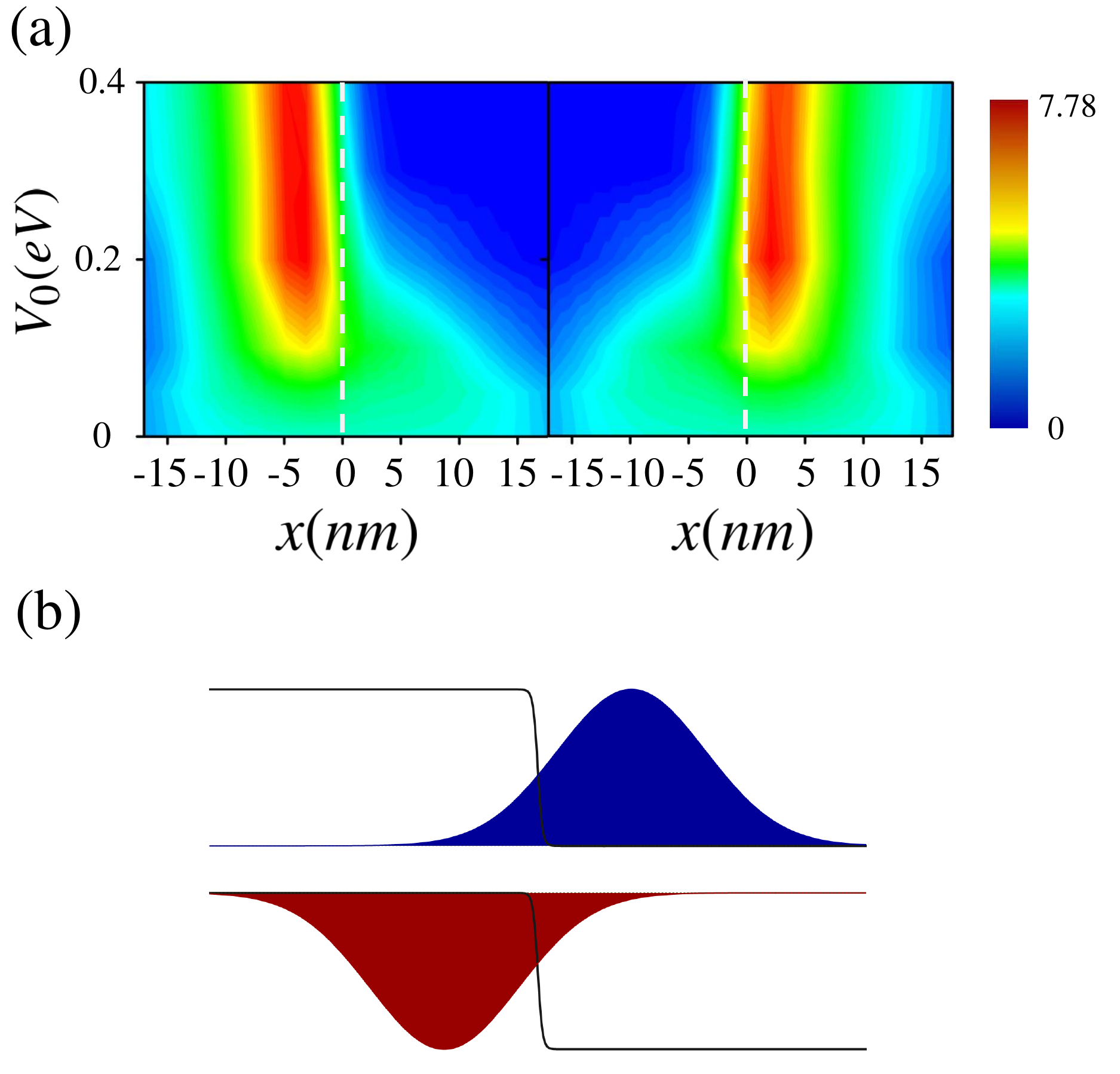}

\protect\protect\protect\protect\caption{(Color online) (a) Left: hole, and right: electron reduced wavefunction 
versus $V_{0}$. Since the translational symmetry is perseved along $y-$direction, we only show the cross-section of the reduced wavefunction along x-direction at $y=0$. Apparently
the electron-hole pair is effectively separated at high $V_{0}$.
However, for $V_{0}\sim0.1-0.2$ eV a finite tail into the barrier
region remains which give rise to a considerable electron-hole overlap.
(b) Schematics of an interface exciton for $V_{0}>0.1$ eV.  A considerable overlap between the electron and hole wavefunctions survives even for a great spatial separation of the electron and hole. Here, the thin lines denote the band offsets. The dark blue and dark red regions respectively denote the cross-section of the electron and hole's reduced wavefunctions along x-direction at $y=0$.}

\label{fig:fig4}
\end{figure}


We obtained an interface exciton binding energy of about $0.2$ eV
which is of the same order as a 2D exciton in TMDs. Such a large binding
energy at a type-II interface is not present in most conventional
semiconductor nanostructures \cite{Bastard88,Rorison93,Warnock93}.
In fact, a type-II interface exciton is often considered unstable
in conventional semiconductor heterostructures \cite{Bastard88,Degani90}
unless in the presence of other physical structures like a E-field
\cite{Degani90} or within a quantum dot \cite{Rorison93}. In a TMD
lateral heterojunction, however, with a relatively large binding energy a type-II
interface exciton is predicted to be stable with our calculations.
It is also of concern whether the interface exciton changes back to
a 2D exciton easily. From Fig.~\ref{fig:fig2}(b) we see that at
$V_{0}=0.2eV$ the binding energy of the interface exciton is about $0.22eV$,
which is about $0.1eV$ smaller than the binding energy of 2D exciton.
In this sense, we may assume that for realistic configurations the interface
exciton is a stable ground state of a lateral heterojunctions. The
relatively large binding energy exactly results from the weaker screening
effect of Coulomb interaction when the electron-hole separation increases
as shown in Eq.(\ref{eq:Veff}).

It is noted that for a large $V_{0}$ the on-site Coulomb interaction
for an interface exciton is irrelevant. This implies that the electron
and hole are well-separated into opposite regions for sufficiently
large $V_{0}$, while for small and intermediate $V_{0}$, even though
qualitative behaviours are similar, numerical values obtained with
different $U$ are quite different. Without loss of generality, $U=-0.79$
eV is assumed in the remainder of the paper as it gives the closest
free exciton $E_{b}$ with Ref. {[}50{]} for 2D exciton.

The center-of-mass part $\Psi_{g}\left(X\right)$ and relative
motion wavefunction with fixed electron position are respectively
depicted in Fig. \ref{fig:fig3}(a) and (b) for different $V_{0}$.
From Fig. \ref{fig:fig3}(a) we see how $\Psi_{g}\left(X\right)$
varies from $V_{0}=0$ to $0.4$ eV. For small band offset
$V_{0}<0.1$ eV $\Psi_{g}\left(X\right)$ is widespread across the
supercell. This is expected since for small $V_{0}$ the electron-hole
pair behaves as a 2D exciton. But for sufficiently high $V_{0}$,
$\Psi_{g}\left(X\right)$ is localized around the interface at $X=0$.
The center-of-mass part of the wavefunction undergoes a transition
from a plane wave to a localized state as $V_{0}$ increases, demonstrating the competition between the Coulomb interaction and the band offset. From Fig.
\ref{fig:fig3}(b), the biased relative motion wavefucntion for large
$V_{0}$ implies that the electron-hole pair tends to be separated
apart well and thus electron-hole overlap is greatly reduced. To further
demonstrate the separation nature of the type-II interface exciton,
the reduced wavefunction of the electron and hole versus $V_{0}$
is shown in Fig. \ref{fig:fig4}(a), where obviously the electron
preferentially stays at left hand side of the interface while the
the hole stays at the right hand side of the interface.

There is still a considerable optical dipole because of the tunnelling
tail of the electron and hole reduced wavefunction. We find that a
small but notable electron-hole overlap still survives. Such electron-hole
overlap can be schematically demonstrated by Fig. \ref{fig:fig4}(b).
The finite magnitude of overlap for $V_{0}>0.1$ eV implies that the
optoelectric properties can be still detected for interface exciton.
From Fig. \ref{fig:fig2} (d) we see that at $V_{0}=0.2$ eV $D$
only drops by half that of 2D excitons, and by one order of magnitude
at $0.3$ eV. Thus at such $V_{0}$ interface exciton still can be exicted by the pumping light. On the other hand, reduction of $D$ suggest a longer lifetime. For very large $V_{0}$, $D$ is a few orders
smaller than 2D excitons, meaning that interface exciton may have
a lifetime far exceeding 2D excitons.

\begin{figure}[ptb]
\includegraphics[width=3.5in]{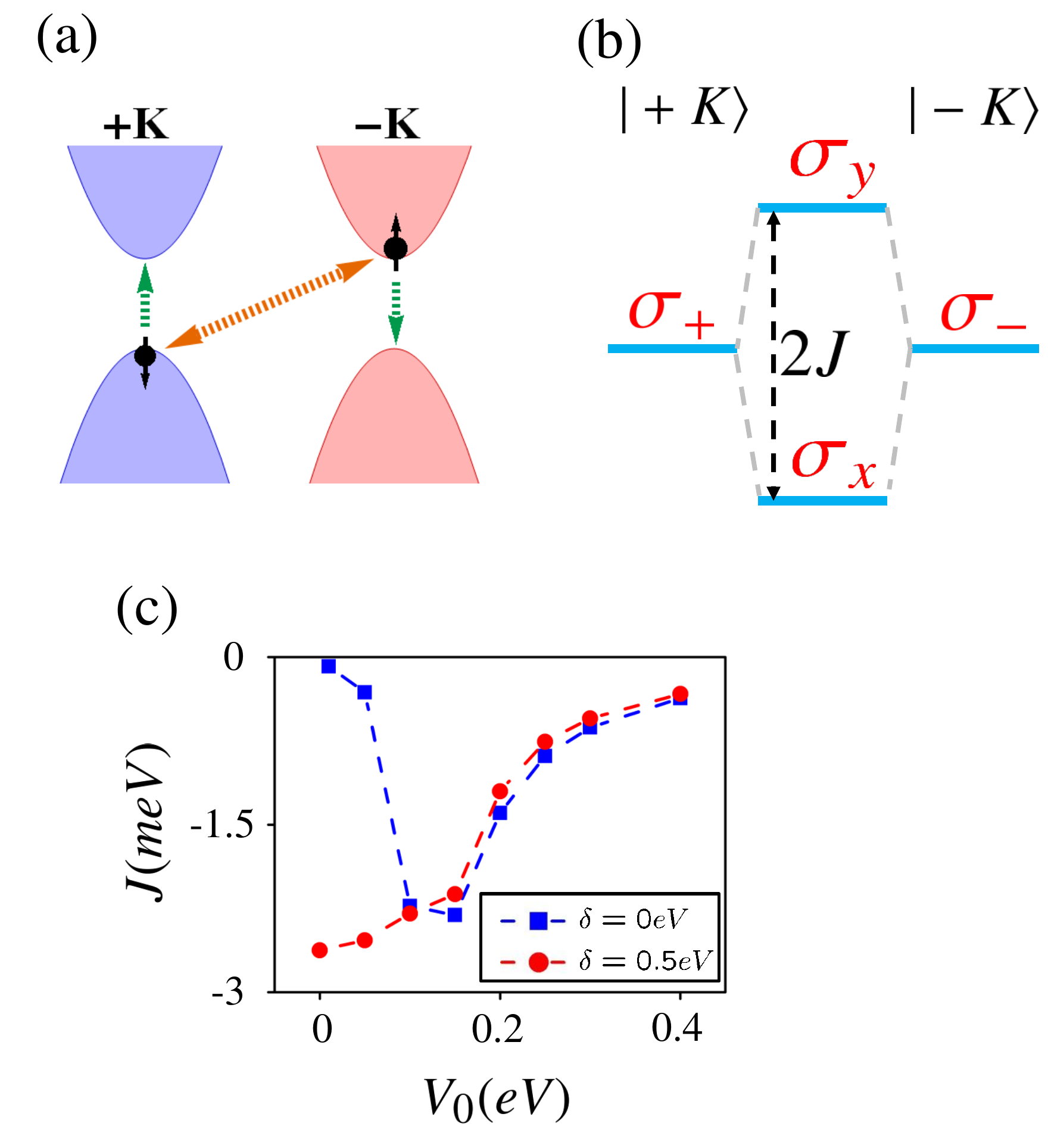}
\caption{(Color online) (a) Illustration of the inter-valley exchange interaction.
Mediated by exchange part of the Coulomb interaction (the orange arrow), the exciton
may change their pesudospin from $-K$ to $+K$ as indicated by the
green arrow, effectively resulting in a valley exchange channel. (b)
The exchange interaction $J$ leads to a splitting between the degenerate
$\pm K$ states. The $\sigma_+,\sigma_-,\sigma_x,\sigma_y$ respectively denotes the polarizations of the light fields
which can pump the corresponding states. The $x$ and $y$ directions are shown in Fig.~\ref{fig:fig1} (a).   (c) The valley coupling strength $J$ versus $V_{0}$ for a symmetric and ansymmetric heterojunction
respectively. Rotational symmetry requires vanishing $J$ at $V_{0}=0$
for a symmetric heterojunction, in contrast to an asymmetric heterojunction
for which $J$ increases at low $V_{0}$. At high voltage valley coupling
of an interface exciton is small regardless of symmetry of the heterojunction because of the reduced electron-hole overlap.}

\label{fig:fig5}
\end{figure}


Finally we calculated the inter-valley coupling strength $J$ for different $V_{0}$ at $U=-0.79$ eV for both a symmetric interface
with $\delta=0$ and an asymmetric interface with $\delta=0.5$ eV
in Fig. \ref{fig:fig5}. For the symmetric case, it is expected that
$J$ tends to zero for $V_{0}=0$ due to the emergence of 3-fold rotation
symmetry. When $V_{0}$ increases, broken symmetry results in a dramatic
increase of intervalley coupling. However for an asymmetric interface
with $\delta=0.5eV$, the 3-fold rotation symmetry is broken at the
beginning and thus there is a considerable $J$ at $V_{0}=0.$ However
as $V_{0}$ further increases, the broken symmetry plays a minor role
and a very similar monotonic decreasing behavior in $J$ is observed
for both interfaces. This manifests the reduced electron-hole overlap
$\psi(P_{X})$ as in the drop of $D$ when $V_{0}$ increases.

A non-zero inter-valley coupling between $\pm K$ implies that interface
exciton ground state has a valley part of the form $\frac{1}{\sqrt{2}}\left(\vert K\rangle\pm\vert-K\rangle\right)$.
This suggests that interface exciton couples with linearly polarized
light instead of circularly polarized light as in 2D excitons which is shown in Fig.~\ref{fig:fig5}(b).
Our calculations shows that $J$ has an order of a few meV in Fig.~\ref{fig:fig5}(c).

Under current parameters, we have numerically evaluated the first
and second order terms of a more rigorous Generalized BOA \cite{Kendricka02,Sun90}
and find that even within the intermediate regime of $V_{0}$, the
corrections terms are in the order $10^{-6}$ eV, which is much smaller
than the energy level spacing in the order of $10^{-1}$ eV. Thus
the correction terms may safely be neglected and the zeroth-order
BOA is sufficient for current circumstance.

\subsection{Numerical Results Based on Continuous Model}

\begin{figure}[ptb]
\includegraphics[width=3.2in]{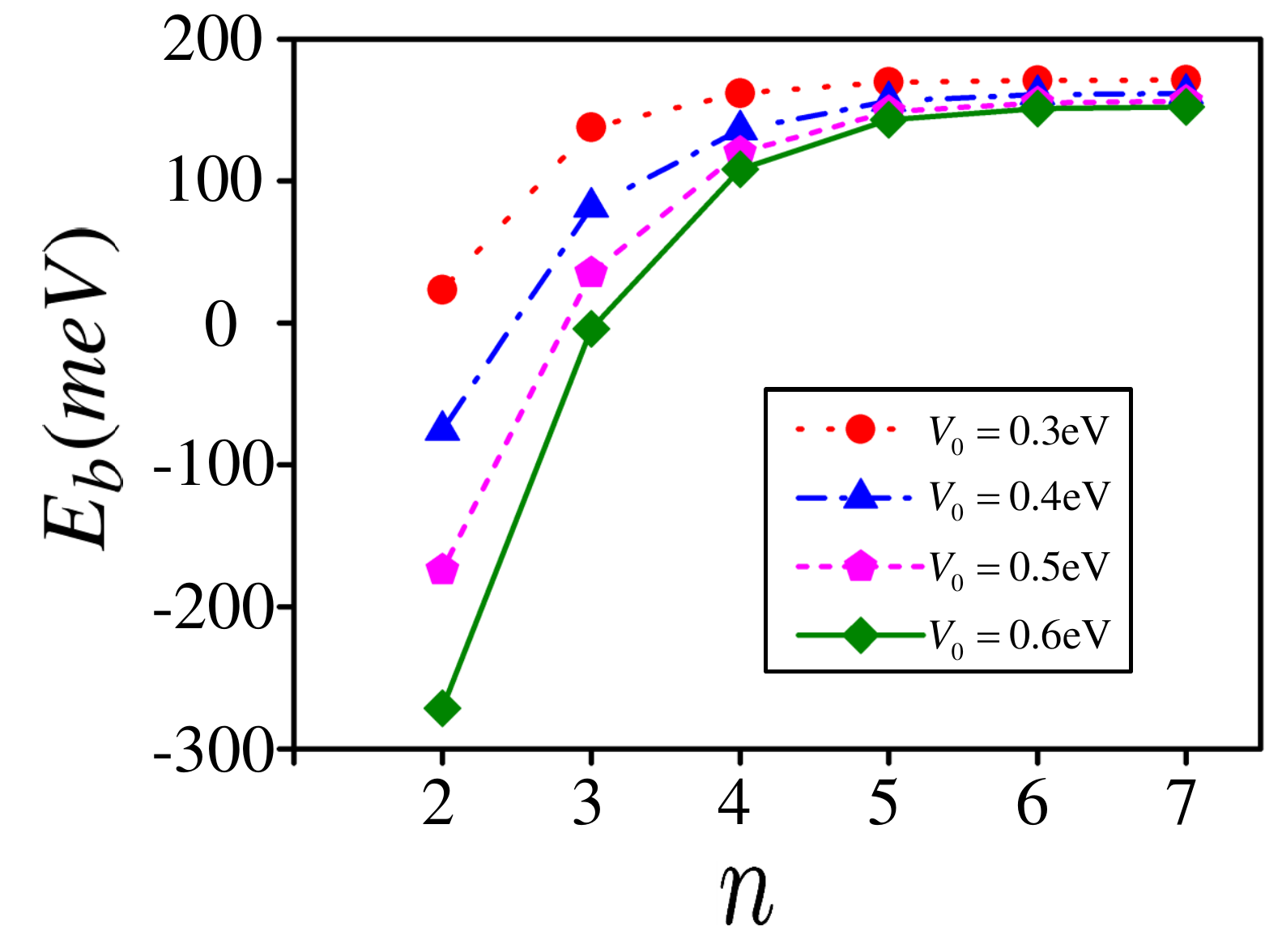}

\caption{(Color online) The binding energy $E_{b}$ versus the cutoff
of principal quantum number $n$ for different potential strength
$V_{0}$. The dotted red line
with circle symbol, dot-dashed blue line with triangle symbol, magenta
short-dashed line with pentagon symbol and olive solid line with diamond
symbol represent the ground state energies with potential strength
$V_{0}=0.3,0.4,0.5,0.6$ eV, respectively. The other parameters
are chosen as $\varepsilon\approx1.1\varepsilon_{0},m_{e}=0.434m_{0},$ and $m_{h}=0.533m_{0}$ with vacuum dielectric constant $\varepsilon_{0}$ and free electron mass $m_{0}$. It is clear that the ground state energies converge
very quickly along with increasing principal quantum number even for
relative large potential strength.}

\label{fig:fig6}
\end{figure}


When the size of the supercell is much larger than the lattice constant,
we can also introduce a continuous model of type II interface exciton,
where its Hamiltonian in Eq. (\ref{eq:4}) is diagonalized with a
2D hydrogenic basis. In the Hilbert space expanded by the 2D hydrogenic
basis $\{\phi_{nl}\left(r\right)\}(n=0,1,\ldots,l=-n+1,\ldots,n-1)$
which satisfy the Schrödinger equation of the usual 2D hydrogen atom\cite{Yang91}
\begin{equation}
\left(-\frac{\hbar^{2}}{2\mu}\nabla_{\mathbf{r}}^{2}-\frac{e^{2}}{\varepsilon r}\right)\phi_{nl}\left(r\right)=E_{n}\phi_{nl}\left(r\right),
\end{equation}
Eq. (\ref{eq:4}) can be rewritten as
\begin{equation}
a_{nl}^{k}\left(X\right)\left[E\left(X\right)-E_{k}\left(X\right)\right]+\sum_{n'=1}^{\infty}\sum_{l'=-n+1}^{n-1}V_{n'l'}^{nl}\left(X\right)a_{n'l'}^{k}\left(X\right)=0,
\end{equation}
where $\Theta_{k}\left(X,\mathbf{r}\right)=\sum_{n,l}a_{nl}^{k}\left(X\right)\phi_{nl}\left(\mathbf{r}\right)$
has already been assumed as the linear combination of the basis with
coefficients $\{a_{nl}^{k}\left(X\right)\}$ and the elements of the
electric potential are defined as $V_{n'l'}^{nl}\left(X\right)=\int d\mathbf{r}\phi_{nl}^{*}\left(\mathbf{r}\right)\left[V_{\mathrm{C}}\left(r\right)+V_{\mathrm{I}}\left(X,\mathbf{r}\right)\right]\phi_{n'l'}\left(\mathbf{r}\right).$

\begin{figure}[tp]
\includegraphics[width=3.5in]{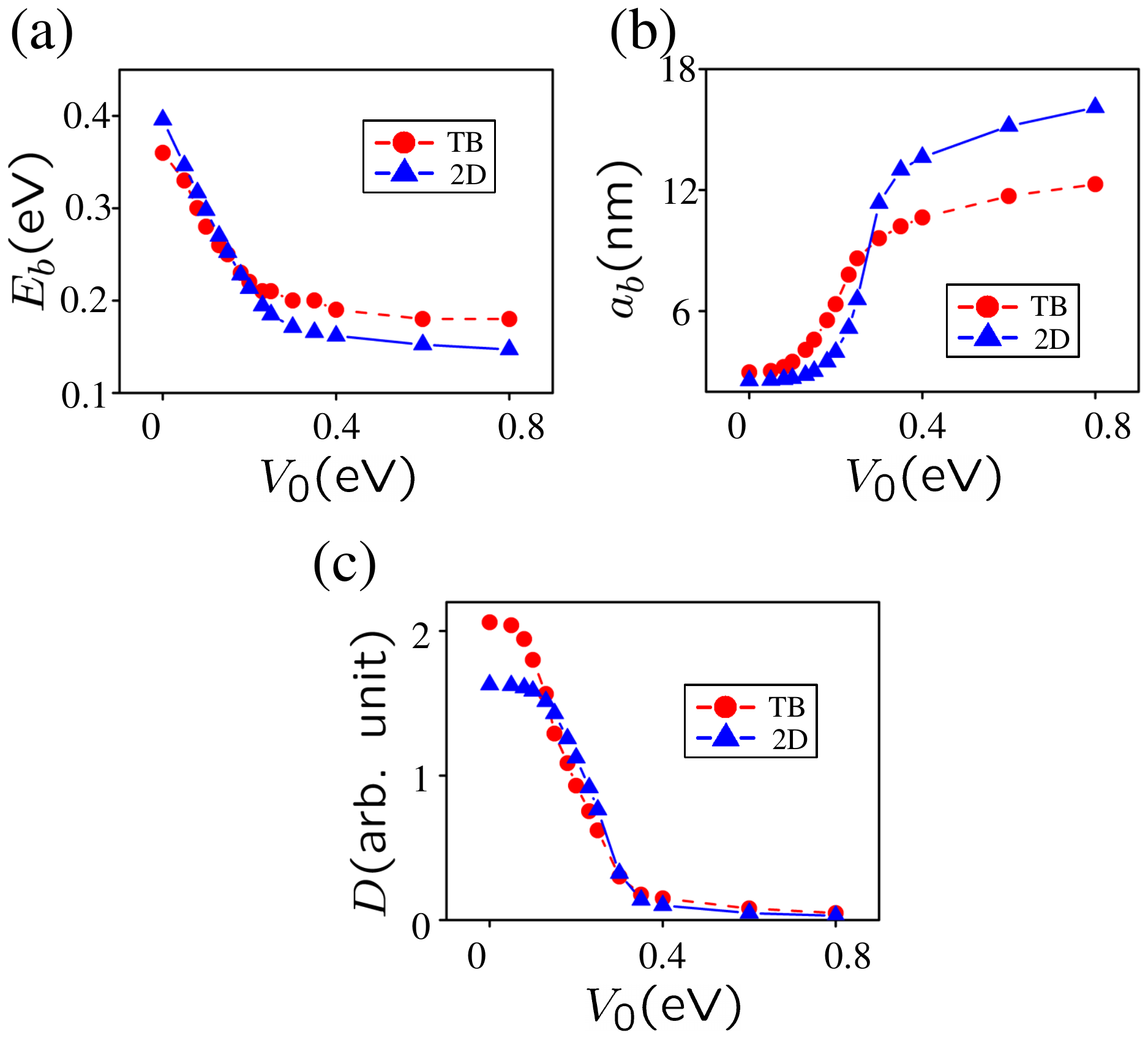}

\caption{(a) The binding energy $E_{b}$,
(b) the exciton radius $a_{b}$, and (c) the optical dipole $D$ (in arbitrary
unit) versus the strength of the band offset $V_{0}$. The
blue triangle symbols with solid line and the red sphere symbols with
dashed line respectively represent the numerical results obtained
from TB model and continuous model, which are respectively denoted
as \char`\"{}TB\char`\"{} and \char`\"{}2D\char`\"{} in the plot.}

\label{fig:fig7}
\end{figure}

Since we need an infinite principal quantum number $n$ to complete
the Hilbert space of Eq. (\ref{eq:4}) which is clearly impossible,
we need to set a cutoff $n$ when both the binding energy and wavefunction
of the ground state interface exciton are convergent. We plot the binding
energy versus the principal number $n$ for different potential strength
$V_{0}$ in Fig. (\ref{fig:fig6}). It is obvious that the ground
state energies converge very quickly along with the principal quantum
number even for relatively large potential strength. In the following
discussion, the cutoff of $n$ is set as $n_{cutoff}=7.$ It is also
important to note that the dielectric constant $\varepsilon\approx1.10\varepsilon_{0}$
for the 2D hydrogenic basis is fixed in the above numerical calculation
in order to obtain the same binding energy $E_{b}\approx220meV$ as
the one from TB model at $V_{0}=0.2$ eV. Here, $\varepsilon_{0}$ is the vacuum dielectric constant. In this sense, the binding energy of the 2D exciton is  $E_{b}=396meV$. The other parameters are $m_{e}=0.434m_{0}, m_{h}=0.533m_{0}$ with free electron mass $m_{0}$. Based on the continuous
model the numerical results of the binding energy $E_{b}$, effective radius $a_{b}$ and optical dipole
$D$ obtained with different strength of the band offset $V_{0}$
are shown in Fig. (\ref{fig:fig7}) as the blue triangle symbols and
solid lines. Here, the numerical results based on the TB model is
also shown in the same figure as the red sphere symbols and dashed
lines. The numerical results especially the energies resembles each
other reasonably, which implies the validity of both methods. For
large band offset $V_{0}$,  the difference between the numerical results
of both methods becomes greater because the size of the supercell we chose is not sufficiently large.

\subsection{Interface exciton at lateral double heterojunctions}

The former discussion focuses on the properties of the 1-D interface exciton at single heterojunction as shown in Fig.~\ref{fig:fig1}(a). Another important case is the lateraldouble heterojunctions as shown in Fig. ~\ref{fig:fig1}(d). When the interface exciton is generated in such structure, the electron is supposed to locate at the central region and the hole is supposed to locate at both hands side of the central region due to the lattice potential. However, because of the Coulomb interaction the electron and the hole have tendency to bind each other. Such competition will affect the properties of the interface exciton greatly. Since the lattice potential depends on the width of the double heterojunctions as well as the potential strength now, we calculate the binding energy of the interface exciton versus the width of the double heterojunctions $L$  and the potential strength $V_0$ by applying the TB method.
\begin{figure}[ptb]
\includegraphics[width=3.5in]{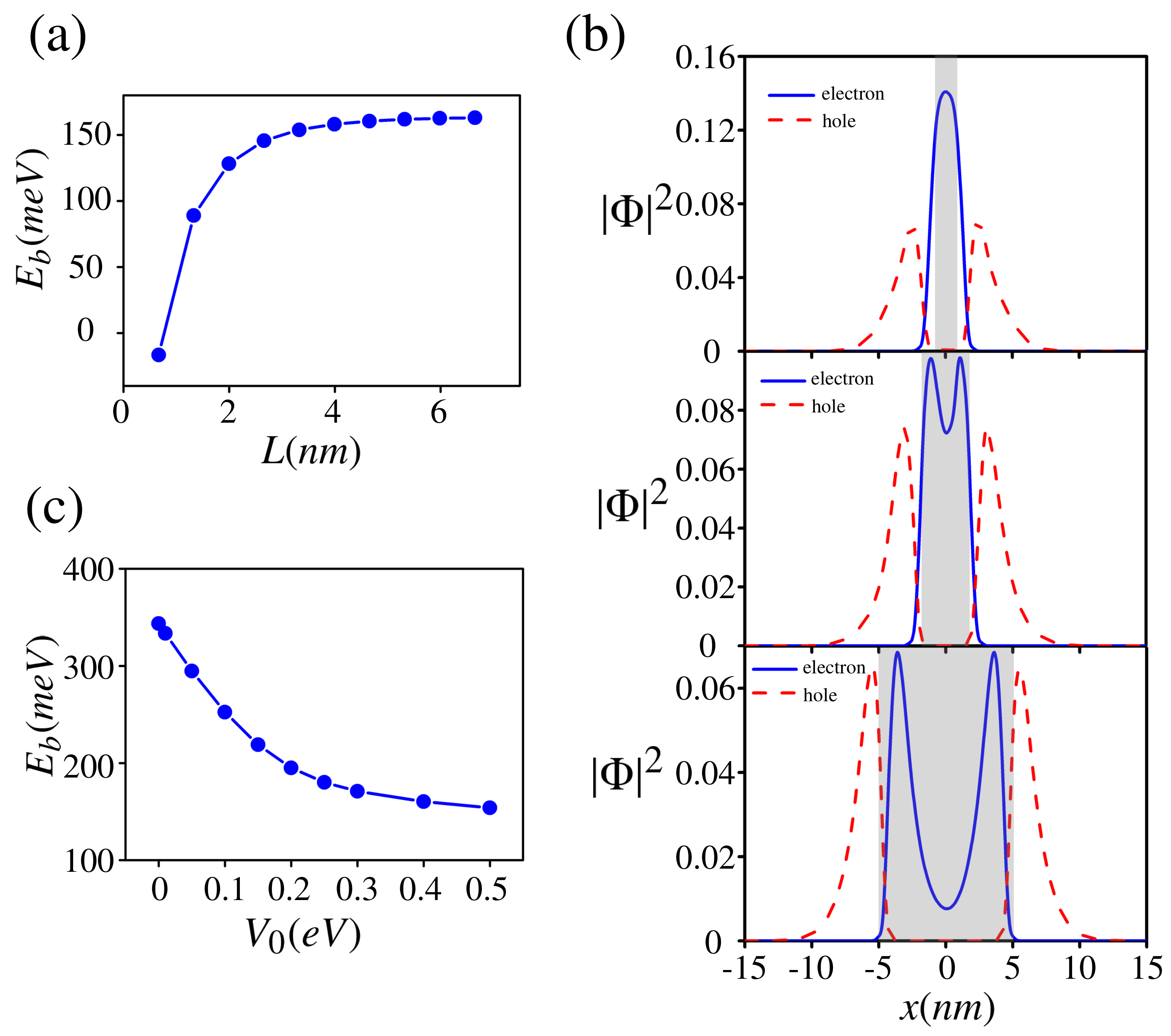}

\caption{(Color online) (a)The binding energy of the interface exciton at  lateral double heterojunctions versus the width of the double heterojunctions $L$. (b)The typical reduced wavefunction of electrons (blue solid lines) and holes (red dashed lines) for different widths of double heterojunctions. The three different widths are $L\approx1.3nm,3.3nm,10nm$. The corresponding central regions are denoted by the shadow areas. See text for the details.  (c) The binding energy of the interface exciton versus the potential strength of the double heterojunctions $V_0$ for $L\approx3.3nm$.}

\label{fig:fig8}
\end{figure}

The binding energy versus the width of the double heterojunctions $L$ is depicted in  Fig.~\ref{fig:fig8}(a). The $V_{0}$  is chosen as $0.5eV$ in Fig.~\ref{fig:fig8}(a). The other parameters are chosen the same as those in Fig.~\ref{fig:fig2}. As shown in Fig.~\ref{fig:fig8}(a), the binding energy increases as the width of the double heterojunctions increases and eventually saturates to a constant value which is the binding energy of the 1D exciton shown in Fig. ~\ref{fig:fig2}(a). It can be interpreted by the overlap of the 1D excitons locating at both interfaces. The reduced wavefunction of electron and hole for different widths of the double heterojunctions $L$ are respectively depicted as blue solid lines and red dashed lines in Fig.~\ref{fig:fig8}(b). The typical effective radius of the 1D exciton for potential strength $V_{0}=0.5eV$ is around $5nm$ as shown in Fig. ~\ref{fig:fig2}(c). For a small double heterojunctions with $L<5nm$, the lattice potential dominates the binding energy of the interface exciton. The consisting electron in 1D excitons locating at both interfaces has great overlap which results in that the electron can only locates at the very center of the double heterojunctions. When the double heterojunctions width increases to be larger than the typical effective radius of the 1D exciton such as $L>5nm$, the Coulomb interaction becomes dominating and the electron prefers to locate in the vicinity of the each interfaces which actually reduces the overlap of the electrons wavefunction. When the width $L$ is much larger than the effective radius, the overlap tends to zero which results in the saturated value equaling to the binding energy of the 1D exciton.

The binding energy versus the potential strength $V_0$  is depicted in  Fig.~\ref{fig:fig8}(c), where  $L$  is chosen as $1.3nm$. The binding energy decreases when the potential strength $V_0$ increases. The reason is that the effective radius of the 1D interface exciton become larger as the potential strength $V_0$ increases as shown in Fig. ~\ref{fig:fig2}(b). Therefore the overlap of the electron wavefunction becomes smaller and eventually reduce the binding energy.

\begin{figure}[ptb]
\includegraphics[width=3.5in]{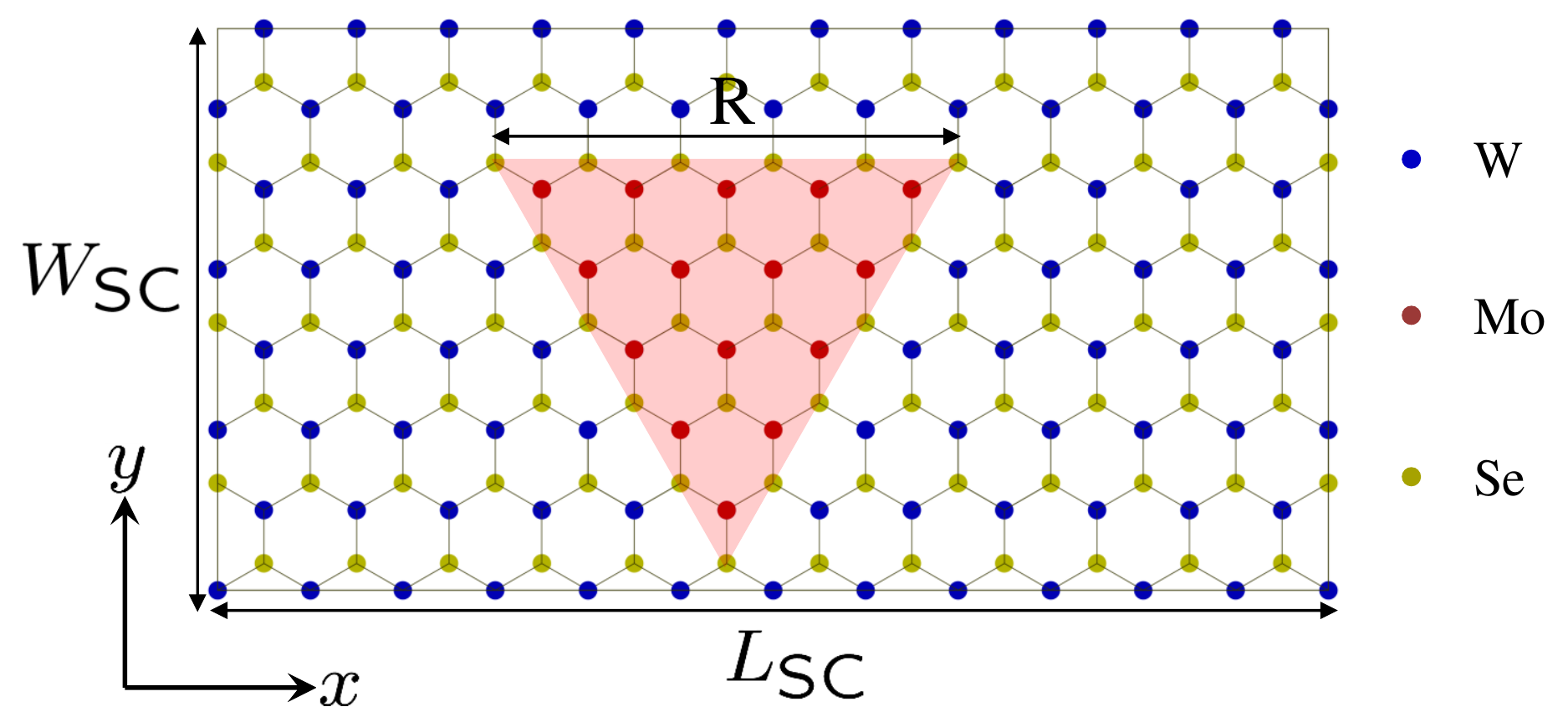}

\caption{(Color online) Schematics of the triangular heterostructure formed
by monolayer $\mathrm{WSe}_{2}-\mathrm{MoSe}_{2}$. The blue, red
and yellow spheres respectively denotes $\mathrm{W}$ atoms, $\mathrm{Mo}$
atoms and $\mathrm{Se}$ atoms respectively. The red shadow region
denotes the region of triangular band offset.}

\label{fig:fig9}
\end{figure}

\section{Interface exciton at closed triangular sharp interface and effective quantum dot confinement}
\subsection{Numerical results of closed triangular sharp interface without valley index}

For all the discussion above, we have assumed the quasi 1D heterojunction
as shown in Fig. (\ref{fig:fig1}). However, the realistic lateral
heterostructures for TMDs present the triangular shape\cite{Duan14,Gong15,Gong14,Huang14},
whose characteristic length scale is about 5$\mu m$. Usually, the
electron-hole separation of interface exciton is up to 10$nm$ for
large $V_{0}$ from the above calculation of 1D interface. It is much
smaller than the characteristic length scale of triangular heterostructures,
which means the calculation of 1D interface is also valid for the
closed triangular sharp interface in current experiments.

If the characteristic length scale of closed triangular sharp interface
decreases to the same order of the electron-hole separation of interface
exciton, the electron (hole) wavefunction will strongly affected by
the boundaries of the triangular shape and thus such quantum confinement effect should be taken into consideration. Actually, such closed triangular interface effectively realizes  0D quantum dot confinement of the interface excitons. From the similar Hamiltonian
in Eq. (\ref{eq:two-body}) but with triangular band offset
as shown in Fig.~(\ref{fig:fig9}), which reads

\begin{equation}
V_{\mathrm{e}}(\mathbf{r})=\begin{cases}
V_{0}, & \mathbf{r}\in\mbox{triangular quantum dot,}\\
0, & \mathbf{r}\notin\mbox{triangular quantum dot,}
\end{cases}
\end{equation}
\begin{equation}
V_{\mathrm{h}}(\mathbf{r})=\begin{cases}
0, & \mathbf{r}\in\mbox{triangular quantum dot,}\\
V_{0}, & \mathbf{r}\notin\mbox{triangular quantum dot.}
\end{cases}
\end{equation}
Here, $L_{\mathrm{SC}}$ and $W_{\mathrm{SC}}$ are the length and
the width of the supercell adopted in the calculations in units of lattice
constant $a$, and $R$ is the edge length of the regular triangular quantum dot.

Since the translational symmetry is no longer preserved in such closed triangular sharp interface, we need to develop another numerical method to calculate
the physical properties of interface exciton. The complete orthonormal
basis $\{\phi_{e}^{(n)}\left(\mathbf{r}_{e}\right)\otimes\phi_{h}^{(n,m)}\left(\mathbf{r}_{h}\right)\}$
are introduced to expand the original Hamiltonian, where $\phi_{e}^{(n)}\left(\mathbf{r}_{e}\right)$
is the n-th eigen-state of the electron confined in the triangular
region without hole part such as
\begin{equation}
\left[-\frac{\hbar^{2}}{2m_{e}}\nabla_{\mathbf{r}_{e}}^{2}+V_{\mathrm{e}}(\mathbf{r}_{e})\right]\phi_{e}^{(n)}\left(\mathbf{r}_{e}\right)=E_{e}^{(n)}\phi_{e}^{(n)}\left(\mathbf{r}_{e}\right),
\end{equation}
and the $\phi_{h}^{(n,m)}\left(\mathbf{r}_{h}\right)$ is the $m-$th
eigen-state of the hole effective Hamiltonian $H_{eff}\left(\mathbf{r}_{h}\right)\phi_{h}^{(n,m)}\left(\mathbf{r}_{h}\right)=E_{h}^{(n,m)}\phi_{h}^{(n,m)}\left(\mathbf{r}_{h}\right)$, 
where the effective Hamiltonian of hole is obtained by averaging the
original Hamiltonian on $\phi_{e}^{(n)}\left(\mathbf{r}_{e}\right)$
as
\begin{equation}
H_{eff}\left(\mathbf{r}_{h}\right)=\int d\mathbf{r}_{e}\phi_{e}^{(n),*}\left(\mathbf{r}_{e}\right)H\phi_{e}^{(n)}\left(\mathbf{r}_{e}\right).
\end{equation}

The Hamiltonian matrix elements are straightforwardly calculated as
\begin{align}
H_{n,m}^{n',m'} & \equiv\int d\mathbf{r}_{e}\int d\mathbf{r}_{h}\phi_{e}^{(n'),*}\left(\mathbf{r}_{e}\right)\phi_{h}^{(n',m'),*}\left(\mathbf{r}_{h}\right)H\times\nonumber \\
 & \phi_{e}^{(n)}\left(\mathbf{r}_{e}\right)\phi_{h}^{(n,m)}\left(\mathbf{r}_{h}\right),
\end{align}
which can be simplified according to the orthogonality of the basis
as
\begin{equation}
H_{n,m}^{n',m'}=\begin{cases}
E_{e}^{(n)}+E_{h}^{(n,m)}, & \mbox{if \ensuremath{n=n'}}\mbox{and \ensuremath{m=m',}}\\
0, & \mbox{if \ensuremath{n=n'}}\mbox{and \ensuremath{m\neq m',}}\\
V_{C}\left(n,n',m,m'\right), & \mbox{if \ensuremath{n\neq n',}}
\end{cases}
\end{equation}
with
\begin{eqnarray}
V_{C}\left(n,n',m,m'\right) & \equiv & \int d\mathbf{r}_{e}\int d\mathbf{r}_{h}\phi_{e}^{(n'),*}\left(\mathbf{r}_{e}\right)\phi_{h}^{(n',m'),*}\left(\mathbf{r}_{h}\right)\times\nonumber \\
 &  & V_{\mathrm{C}}\left(\left|\mathbf{r}_{e}-\mathbf{r}_{h}\right|\right)\phi_{e}^{(n)}\left(\mathbf{r}_{e}\right)\phi_{h}^{(n,m)}\left(\mathbf{r}_{h}\right).
\end{eqnarray}

\begin{figure}[tp]
\includegraphics[width=3.5in]{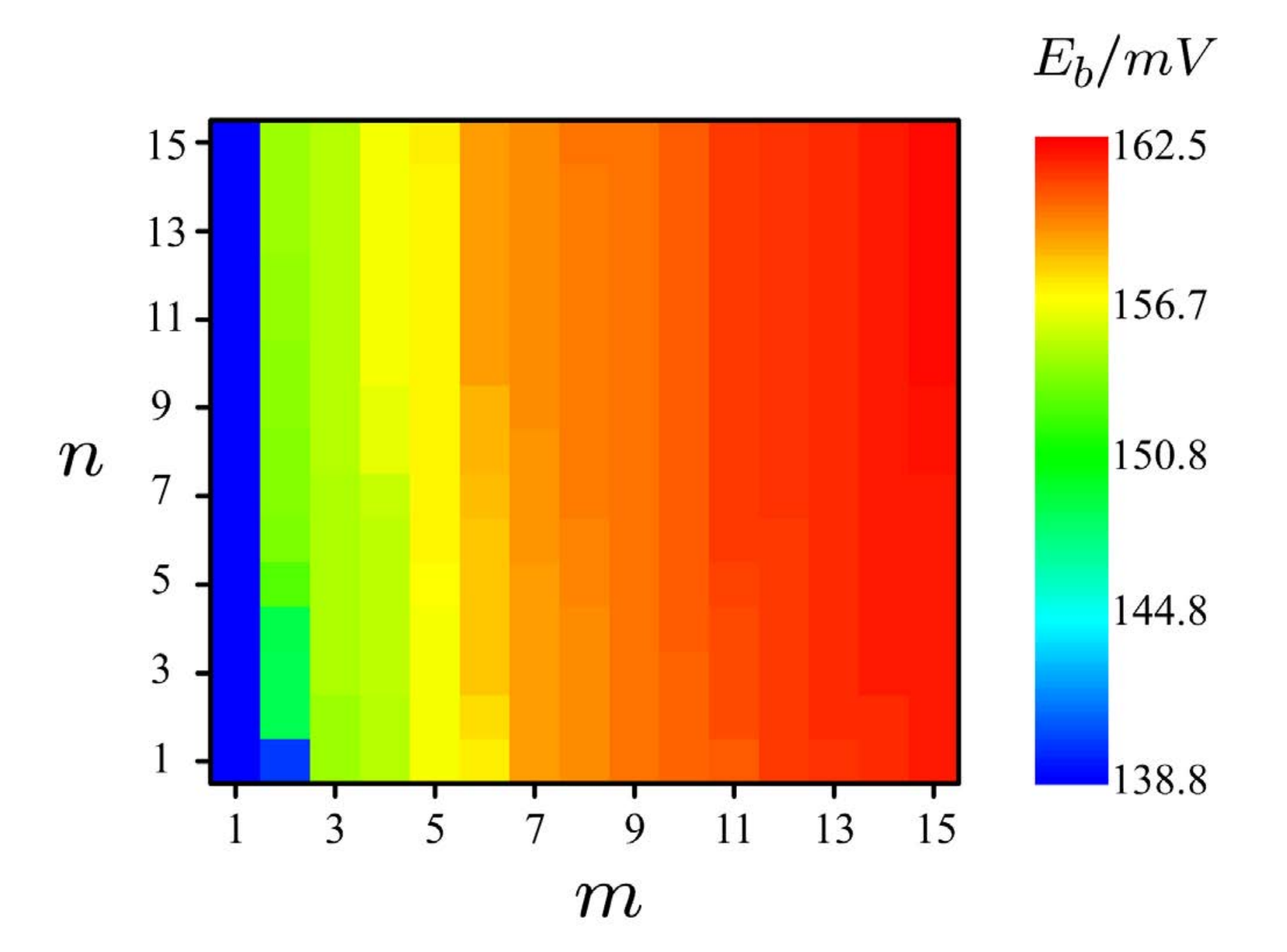}

\caption{(Color online) The binding energy of the exciton of the cloased triangular interface $E_{b}$ versus the quantum number $m$ and $n$. The parameters
are chosen as $R=30a,L_{\mathrm{SC}}=60a,W_{\mathrm{SC}}=36\sqrt{3}a$.
The binding energy converges quickly along the quantum numbers.}

\label{fig:fig10}
\end{figure}

We solve the eigen problem by diagonalizing the Hamiltonian matrix.
We still need to set a cutoff for $n$ and $m$ when the binding energy
of the ground state interface exciton are convergent. As shown in Fig.~\ref{fig:fig10},
where the parameters are chosen as $R=30a,L_{\mathrm{SC}}=60a,W_{\mathrm{SC}}=36\sqrt{3}a$,
clearly the binding energy converges quickly along the quantum numbers,
especially along $n$. In the following calculation, we set the cutoff
$n_{cutoff}=m_{cutoff}=15$.

\begin{figure}[tp]
\includegraphics[width=3.5in]{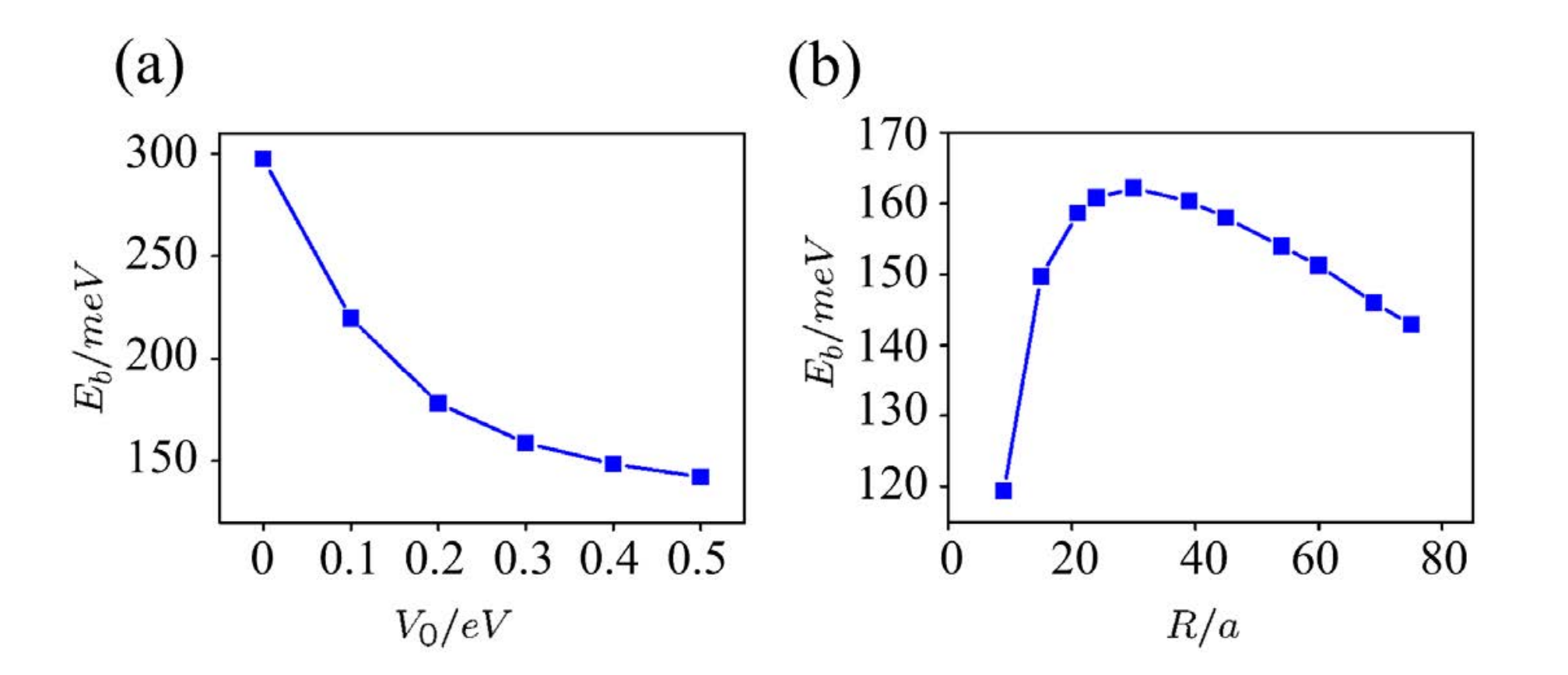}

\caption{(Color online) The binding energy versus (a) the band offset
$V_{0}$ and (b) the size of the quantum dot $R$. The
parameters for (a) are chosen as $R=21a$. The potential strength
is chosen as $V_{0}=0.3eV$. The size of the supercell are sufficiently
large.}

\label{fig:fig11}
\end{figure}


The numerical results of the binding energy versus the band offset
$V_{0}$ and the size of the quantum dot $R$ is depicted
in Fig.~\ref{fig:fig11}. The binding energy monotonically decreases
as the band offset increases as shown in Fig.~\ref{fig:fig11}(a),
which results from the stronger quantum confinement. However, as shown
in Fig.~\ref{fig:fig11}(b), the behavior of the binding energy versus
$R$ has a maximum value due to the competition between the quantum
confinement and the Coulomb interaction. When $R<30a$, basically
the ground state of the electron and hole dominates the wavefunction,
and thus when R increases to decrease the quantum confinement, the
binding energy of exciton increases. While when $R>30a,$ the excited
states of electron and hole start to appear in the wavefunction, which
results in the decrement of binding energy. So if such decrement is greater than the increment of binding energy resulting from quantum confinement,
the binding energy of exciton becomes to decrease as the size of quantum dot R
increases. Therefore, there is a maximum binding energy for an optimal R.

\begin{figure}[tp]
\includegraphics[width=3.5in]{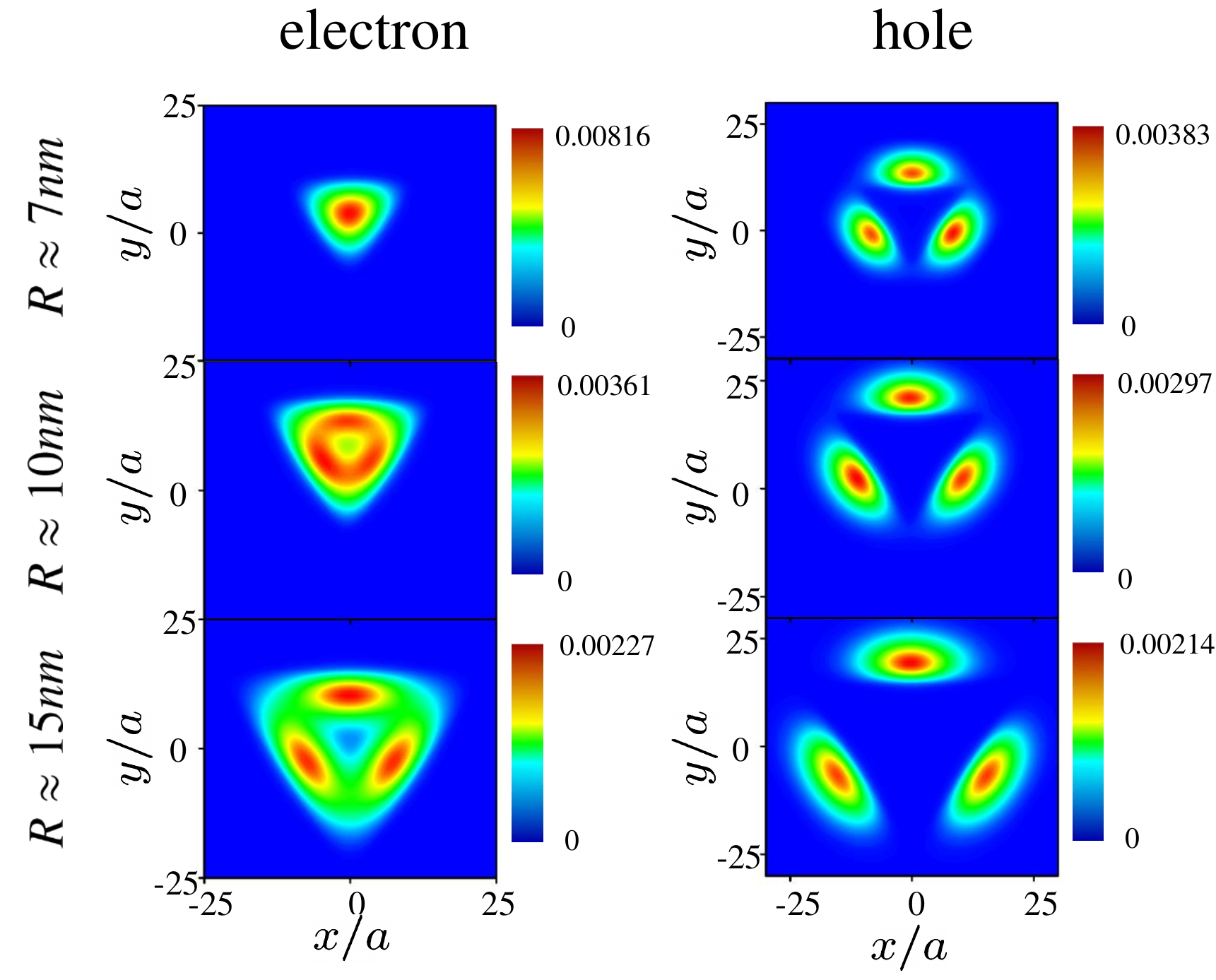}

\caption{(Color online) The reduced wavefunction of the electron (left panel)
and hole (right panel) for closed triangular  interface with different sizes  $R=21a,30a,45a$, namely $R\approx7nm,10nm,15nm$ from the top to bottom. The
potential strength is chosen as $V_{0}=0.3eV$.}

\label{fig:fig12}
\end{figure}

Such competition can be also demonstrated in the reduced wavefunction
of the electron and hole as shown in Fig.~\ref{fig:fig12}. The left
and right panels respectively show the reduced wavefunction of electron
and hole for increasing size of the quantum dot from the
top to bottom. All reduced wavefunctions have the three-fold rotation
symmetry inheriting from the symmetry of the regular triangular shape of the closed interface. For a small quantum dot such as $R=21a,$ the electron are strongly
confined in the quantum dot and the hole wavefunction spreads over
the entire quantum dot. While for a large quantum dot such as $R=45a$,
the wavefunctions of electron and hole only spread over the vicinity
of the edges of the closed interface. Without the interplay between the
wavefunctions at different edges, the closed triangular interface will
degrade to 1D interface case. For a large quantum dot, the binding
energy is about $140meV$, which is consistent with the former 1D
interface calculation as shown in Fig.~\ref{fig:fig7}(b).

We can imagine that when the size of the quantum dot is
much larger than the effective radius of the 1D interface exciton, which is
about 5$nm$ according to the previous calculation, the interface exciton
actually is split into three identical parts locating at the edges
of triangular quantum dot and each part is analogy to the quasi-1D
exciton. When the size of the triangular quantum dot is decreased,
the three parts have considerable overlap at the corners of the triangular
quantum dot when the size of the triangular quantum dot decreases. In this sense,
an effective Hamiltonian is introduced to describe such three-fold
rotational symmetric system as
\begin{equation}
H_{eff}=\left[\begin{array}{ccc}
E_{0} & t e^{i\theta} & t e^{-i\theta}\\
t e^{-i\theta} & E_{0} & t e^{i\theta}\\
t e^{i\theta} & t e^{-i\theta} & E_{0}
\end{array}\right]
\end{equation}
with bases $\{\left|\Phi\right\rangle ,C_{3}\left|\Phi\right\rangle ,C_{3}^{2}\left|\Phi\right\rangle \}.$
Here, $\left|\Phi\right\rangle $ is the wavefunction of 1D
interface exciton at one edge, $C_{3}$ and $C_{3}^{2}$ are rotation
operators of three-fold rotational group, $E_{0}$ is the binding
energy and $t e^{i\theta}$ represents the transition between wavefunctions of
1D interface exciton at different edges. In order to satisfy the three-fold rotation symmetry, the phase factor can
only be $\theta=0,\frac{2\pi}{3}$ or $\frac{4\pi}{3}$. In addition, the phase factors for the opposite valley should be opposite according to the
time-reversal symmetry. Both binding energy $E_{0}$ and
transition coefficient $t$ are determined by the numerical calculation
based on the excitonic lattice model.

\begin{figure}[ptb]
\includegraphics[width=3.4in]{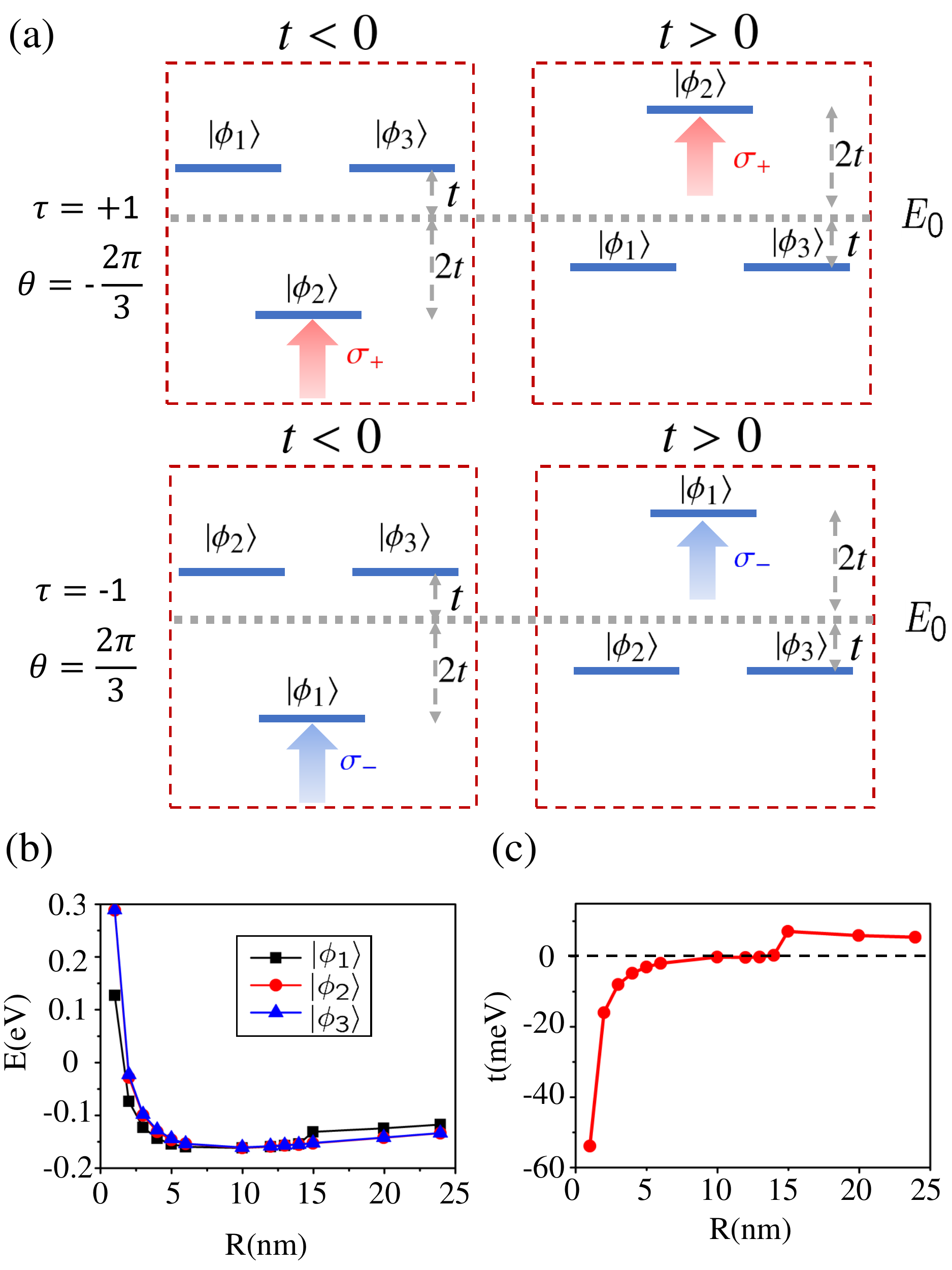}

\caption{(Color online) (a)The energy level schemes for $t<0$ and $t>0$ when $\theta=\mp2\pi/3$ for $\tau=\pm1$ vallyes. The optical selection rules are also depicted. Here, $\tau=\pm1$ are valley index denoting K and -K valleys. The light red and light blue arrows respectively denote the right ($\sigma_+$) and left ($\sigma_-$) polarizations of the light fields coupling to the corresponding exciton states.  (b) The numerical results of the eigen-energies $E$
and (c) the absolute value of the transition coefficient $t$ versus the size of the triangular
quantum dot $R$. Obviously, there is a transition when the
ground state varies from a nondegenerate state to degenerate one.
Additionally, the transition coefficient changes from negative value
to a positive one.}

\label{fig:fig13}
\end{figure}

By diagoalizing the effective Hamiltonian, the lowest three excitonic
states can be found as
\begin{align}
\left|\phi_{1}\right\rangle  & =\frac{1}{\sqrt{3}}\left(e^{i\pi}\left|\Phi\right\rangle +e^{i\frac{\pi}{3}}C_{3}\left|\Phi\right\rangle +e^{i\frac{5\pi}{3}}C_{3}^{2}\left|\Phi\right\rangle \right),\label{eq:21}\\
\left|\phi_{2}\right\rangle  & =\frac{1}{\sqrt{3}}\left(\left|\Phi\right\rangle +e^{i\frac{2\pi}{3}}C_{3}\left|\Phi\right\rangle +e^{i\frac{4\pi}{3}}C_{3}^{2}\left|\Phi\right\rangle \right),\label{eq:22}\\
\left|\phi_{3}\right\rangle  & =\frac{1}{\sqrt{3}}\left(\left|\Phi\right\rangle +C_{3}\left|\Phi\right\rangle +C_{3}^{2}\left|\Phi\right\rangle \right),\label{eq:23}
\end{align}
with corresponding eigen-energies $E_{i}=E_{0}+2t\cos \left(\frac{2i\pi}{3}-\theta \right)$,(i=1,2,3). Since $\theta$ can only be $0,\frac{2\pi}{3}$ or $\frac{4\pi}{3}$, there are two degenerate states. We take $\theta=\mp2 \pi/3$ for $\tau=\pm1$ as an example. The energy level scheme is depicted in Fig.~\ref{fig:fig13}(a), where obviously $\left|\phi_{2}\right\rangle $ and $\left|\phi_{3}\right\rangle $ are degenerate states. More interest fact is that there is a transition when the absolute value of the transition coefficient $t$ varies from negative
value to a positive one. When $t<0$, the $\left|\phi_{1}\right\rangle $
is ground state. In contrast, when $t>0,$ the degenerate states $\left|\phi_{2}\right\rangle $
and $\left|\phi_{3}\right\rangle $ become ground states.

Such transition is depicted in the Fig.~\ref{fig:fig13}(a).
The numerical calculation based on the lattice model show the transition
occurs when the size of the triangular quantum dot is about 12.5nm
(Fig~\ref{fig:fig13}(b)). The transition coefficient $t$ can also
be parameterized by the numerical calculation, which is shown in Fig~\ref{fig:fig13}(c).
The absolute value of transition coefficient $t$ strongly depends on the overlap of
the quasi-1D excitonic wavefunctions at the corners of the triangular
quantum dot. For a small quantum dot, the electron confined in the
quantum dot and thus the overlap of the electron part supplies a relatively
large attractive Coulomb interaction to overcome the kinetic energy.
Therefore $t$ has negative value. In contrast for a large quantum
dot, both the electron and hole spread over the vicinity of the edges
of the quantum dot, and thus the overlap of the electron and hole
are greatly decreased. In this sense, the Coulomb interaction part
becomes smaller than the kinetic part resulting in positive $t$.

Among the above three excitons, only one state is bright exciton and the other two states are dark excitons when pumping them with right ($\sigma_+$) or left ($\sigma_-$) circularly polarized light. The optical transition matrix elements of those excitons are proportional to $\left\langle \phi_{i}\right|P_{\pm}\left|vac\right\rangle, (i=1,2,3)$, where $P_{\pm}$ are the dipole moments corresponding to the $\sigma_+$ or $\sigma_-$ circularly polarized light and $\left|vac\right\rangle$ denotes the initial states with full valence bands and empty conduction bands. Since under the three-fold rotation the transformations of the dipole moments are
\begin{equation}
C_{3}^{-1}P_{\pm}C_{3}=e^{\pm i\frac{2\pi}{3}}P_{\pm},\left(C_{3}^{2}\right)^{-1}P_{\pm}C_{3}^{2}=e^{\pm i\frac{4\pi}{3}}P_{\pm},
\end{equation}
only the exciton states with appropriate phase factors of coefficients have nonzero optical transition matrix elements and thus are bright excitons. In this sense, the $\sigma_-$ circularly polarized light can pump the $\left|\phi_{1}\right\rangle$ in the $\tau=-1$ valley, and  $\sigma_+$ circularly polarized light can pump the $\left|\phi_{2}\right\rangle$ in the $\tau=+1$ valley. The corresponding optical selection rule is shown in the Fig.~\ref{fig:fig13}(a).


\subsection{Numerical results of closed triangular sharp interface with valley index}

In the above discussion, only the interface excitons in the same valley
are discussed. However, due to the geometry of the closed triangular sharp interface, the
inter-valley couplings are inevitable which eventually couple interface
excitons in the opposite valleys. We start from the effective Hamiltonian
both including the three-fold rotational symmetric excitons
and the inter-valley couplings as
\begin{equation}
H_{eff}^{inter}=\left[\begin{array}{cccccc}
E_{0} & te^{i\theta_+} & te^{-i\theta_+} & p & q & q\\
te^{-i\theta_+} & E_{0} & te^{i\theta_+} & q & p & q\\
te^{i\theta_+} & te^{-i\theta_+} & E_{0} & q & q & p\\
p & q & q & E_{0} & te^{i\theta_-} & te^{-i\theta_-}\\
q & p & q & te^{-i\theta_-} & E_{0} & te^{i\theta_-}\\
q & q & p & te^{i\theta_-} & te^{-i\theta_-} & E_{0}
\end{array}\right],\label{eq:24}
\end{equation}
where the bases in the real space $(i=1,2,3)$ are
\begin{equation}
\mbox{\ensuremath{\left\langle \mathbf{r_{e},r_{h}}|\phi_{i},\tau\right\rangle \approx\exp\left(i\tau\mathrm{\mathbf{K}}\cdot\mathrm{\mathbf{\left(r_{e}-r_{h}\right)}}\right)}}\left\langle \mathbf{r_{e},r_{h}}|\phi_{i}\right\rangle \times u_{\mathbf{K}}\left(\mathbf{r_{e},r_{h}}\right)\label{eq:25}
\end{equation}
and $\tau=\pm1$ is the valley index. The $\left|\phi_{i}\right\rangle $
are the envelopes of the excitonic states without considering the
valley index which are defined in Eq.(\ref{eq:21}-\ref{eq:23}) and
$u_{\mathbf{K}}\left(\mathbf{r_{e},r_{h}}\right)$ is the periodic
parts of the Bloch wavefunctions. We adopted the assumption that $u_{\mathbf{K+q}}\left(\mathbf{r_{e},r_{h}}\right)\approx u_{\mathbf{K}}\left(\mathbf{r_{e},r_{h}}\right)$
for the sake of simplicity. Here, $E_{0}=\left\langle \phi_{i},\tau\right|H\left|\phi_{i},\tau\right\rangle $
is the binding energy of the 1D interface exciton, $t=\left\langle \phi_{i},\tau\right|HC_{3}\left|\phi_{i},\tau\right\rangle $
is the intra-valley inter-edge hoppings, $p=\left\langle \phi_{i},\tau\right|H\left|\phi_{i},\overline{\tau}\right\rangle $
is the inter-valley intra-edge hoppings, and $q=\left\langle \phi_{i},\tau\right|HC_{3}\left|\phi_{i},\overline{\tau}\right\rangle $
is the inter-valley inter-edge hoppings. Here, the original Hamiltonian
$H$ is introduced in Eq. (\ref{eq:two-body}). In order to make sure that the ground state of the interface exciton still inherit the same optical selection rule, which means that the $\sigma_+$ ($\sigma_-$) circularly polarized light only pump the ground states of the excitons in the $\tau=+1$ ($\tau=-1$) valley, the phase factors for both valleys are fixed as $\theta_+=-\theta_-=-2\pi/3$.

\begin{figure}[ptb]
\includegraphics[width=3.5in]{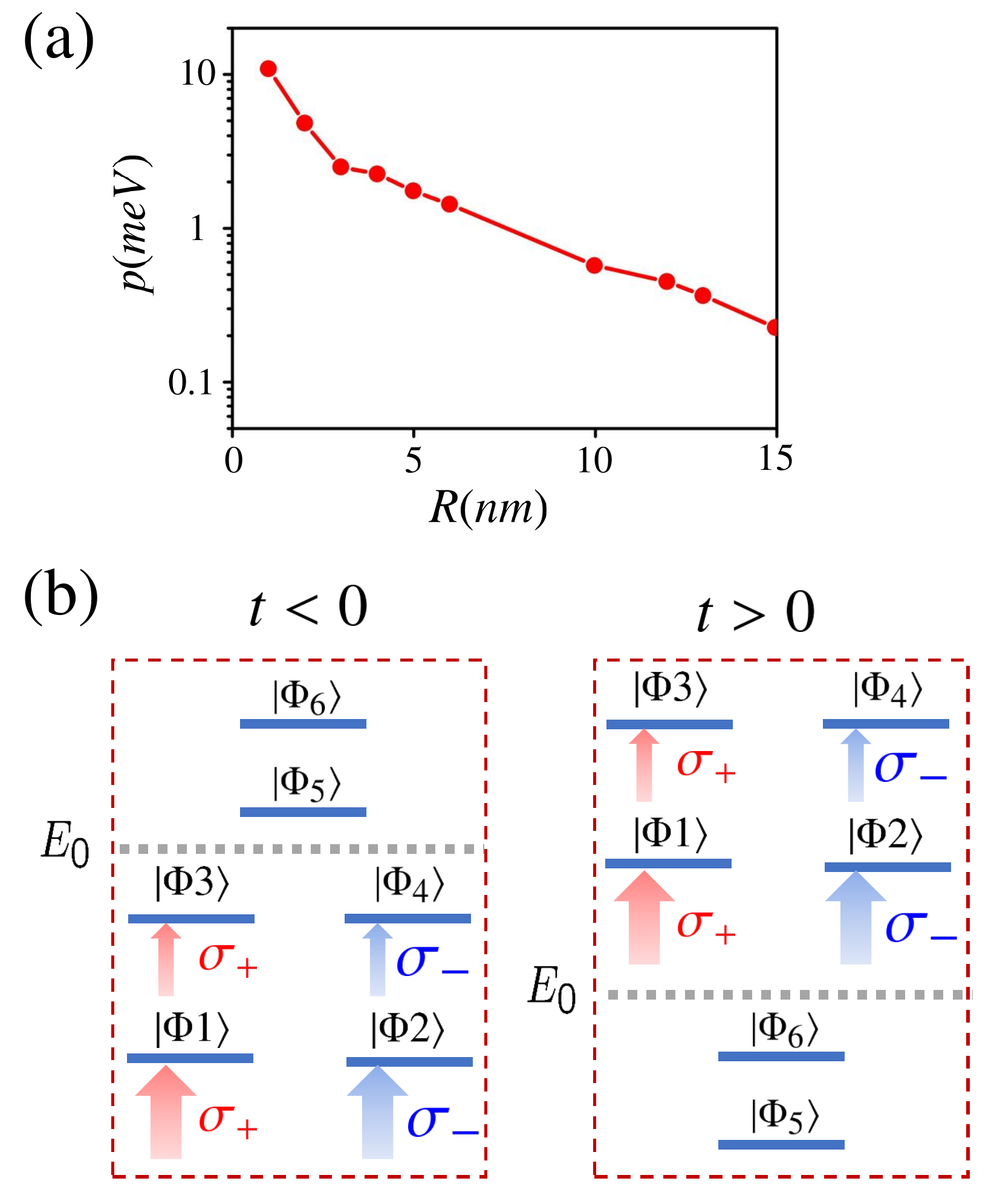}

\caption{(Color online) (a) The inter-valley intra-edge hopping $p$ versus the
size of quantum dot $R$. The magnitude of $p$ almost decrease exponentially
as the size of the quantum dot increases. (b) The energy level schemes for $t<0$ and $t>0$ for $\theta_+=-\theta_-=-2\pi/3$. The optical selection rules are also depicted. The light red and light blue arrows respectively denote the right ($\sigma_+$) and left ($\sigma_-$) polarizations of the light fields coupling to the corresponding exciton states. The different sizes of the arrows denote the coupling strengthes between the exciton states and the light field.}

\label{fig:fig14}
\end{figure}

Since the inter-valley terms $p$$(q)$ are at least one order smaller
than the corresponding intra-valley terms $E_{0}$$(t)$ due to the
large momentum difference, and the inter-valley terms results from
the wavefunction overlap at the corners which obviously become smaller
when the size of the quantum dot increases, the magnitudes of the
parameters have the following relations $E_{0}\gg t\sim p\gg q.$
In the following calculation, we ignore the inter-valley inter-edge
hoppings $q.$ The numerical results of the inter-valley intra-edge
hopping $p$ versus the size of quantum dot $R$ is shown in Fig.~\ref{fig:fig14}.
The magnitude of $p$ almost decrease exponentially as the size of
the quantum dot increases. For the large quantum dot, the value matches the previous inter-valley coupling results shown in
Fig.~\ref{fig:fig5} because the the interface exciton degrades to 1D interface exciton without the the wavefunction overlap at the corners.
 For small quantum dot such as $R<5nm$ it can reach to several meV.

Although $q$ is small in comparison with $t$, it can still couples
the interface excitons in opposite valleys. By diagonalizing the effective
Hamiltonian $H_{eff}^{inter}$ in Eq. (\ref{eq:24}), we can obtain
the lowest six interface excitonic states as
\begin{eqnarray}
\ensuremath{\left|\Phi_{1}\right\rangle } & = & \cos\psi\left|\phi_{2},\tau\right\rangle +\sin\psi\left|\phi_{3},\overline{\tau}\right\rangle ,\\
\left|\Phi_{2}\right\rangle  & = & \cos\psi\left|\phi_{1},\overline{\tau}\right\rangle +\sin\psi\left|\phi_{3},\tau\right\rangle ,\\
\left|\Phi_{3}\right\rangle  & = & -\sin\psi\left|\phi_{2},\tau\right\rangle +\cos\psi\left|\phi_{3},\overline{\tau}\right\rangle ,\\
\left|\Phi_{4}\right\rangle  & = & -\sin\psi\left|\phi_{1},\overline{\tau}\right\rangle +\cos\psi\left|\phi_{3},\tau\right\rangle ,\\
\left|\Phi_{5}\right\rangle  & = & \frac{1}{\sqrt{2}}\left(\left|\phi_{1},\tau\right\rangle -\left|\phi_{2},\overline{\tau}\right\rangle \right),\\
\left|\Phi_{6}\right\rangle  & = & \frac{1}{\sqrt{2}}\left(\left|\phi_{1},\tau\right\rangle +\left|\phi_{2},\overline{\tau}\right\rangle \right),
\end{eqnarray}
with $\tan{\psi}=2p/(3t)$ and the corresponding energies $E_{1}=E_{2}=E_{0}+\frac{1}{2}\left(t-\sqrt{9t^{2}+4p^{2}}\right),E_{3}=E_{4}=E_{0}+\frac{1}{2}\left(t+\sqrt{9t^{2}+4p^{2}}\right),E_{5}=E_{0}-t-p$ and $E_{6}=E_{0}-t+p.$ Since the intervalley coupling $p$ actually couples one bright exciton and one dark exciton such as $\left|\phi_{2},\tau\right\rangle$ and
$\left|\phi_{3},\overline{\tau}\right\rangle$,  both the $\left|\Phi_{1}\right\rangle$ and $\left|\Phi_{3}\right\rangle$ become bright but with different coupling strength with the same right circularly polarized light field. The energy level and the complete optical selection rule are shown in the Fig.~\ref{fig:fig14}(b). According to the orthogonality of the periodic parts of the Bloch
wavefunctions, the envelope wavefunction of the six excitonic states with
the valley index are analogous to the wavefunction without the valley index shown
in Fig.\ref{fig:fig12}.

\section{Conclusion}

In this paper, we theoretically study the interface exciton states
at various lateral heterojunctions of monolayer semiconductors including single, double and closed triangular interfaces. When taking the
distance dependent screening of Coulomb interaction into consideration,
we numerically study the physical observables of type II interface
exciton including the binding energy, effective radius
between the electron and hole and optical dipole. Usually, such problem is quite difficult to be numerically solved by \textit{ab initio} calculations.
We adopted two different approaches to calculate excitons. One approach bases on a
real-space tight binding model, and the other approach considers the
perturbation expansion in a hydrogen-like basis in an effective mass
model. The numerical study shows that even when the electron-hole
separation is much larger compare to the 2D excitons in TMDs, type
II interface exciton still has strong binding energy.
When the effective radius between the electron and the hole is up
to four times of the Bohr radius of 2D excitons, the binding energy remains
1/2 that of 2D excitons. This can be interpreted by the weaker screening
of Coulomb interaction as the electron-hole spatial separation increases.
Large energy separation between interface exciton and 2D excitons for band offset above $0.2$ eV ensures that such 1D interface excitons are stable ones. Due to the spatial indirect nature of type II interface
exciton, exciton radius increases while optical transition dipole
decreases as band offset increases. Still, the optical dipole is
comparable to that of 2D excitons at moderate band offset of 100meV or below. Inter-valley
coupling that arises from electron-hole exchange is also studied,
which may leads to the longitudinal-transverse splitting with the interface breaking the rotational symmetry. The lateral heterojunctions with closed triangular interface is also studied, which realize the 0D quantum dot confinement of exciton. The numerical study shows that the energy level schemes and valley optical selection rules of the exciton in quantum dot depends on the
size of the quantum dot.  Together with valley index, there are more
exciton states in a single quantum dot which can be used to carry
information. With its unique nature of having one carrier confined within the triangle by the band offset and the other carrier bounded to the proximity exterior of the triangle by the strong Coulomb, it is possible to realize the strong excitonic coupling between the neighbouring quantum dots for mediating controlled interplayer between spins at different dots~\cite{Liu10}. In this sense, our investigation may facilitate the quantum information procession based on the 2D monolayer semiconductors.

\appendix

\section{\label{App:gboa}Genralized Born-Oppenheimer approximation}

The regular Born-Oppenheimer approximation only consider the lowest
order of the ratio between the reduced mass and the total mass $\kappa=\mu/M$
which corresponds to the relative motion and the center-of-mass motion,
respectively. The total wavefunction is expressed as a product of
the relative and center-of-mass parts when the adiabatic condition
is satisfied. However, in the problem of our interface exciton there
exist nonadiabatic processes and higher order terms in the ratio $\kappa$
should be taken into consideration. In this sense, we present the
generalized Born-Oppenheimer approximation and the corresponding second-order
perturbation theory here.

The Schrödinger equation satisfied by one center-of-mass motion and
multiple relative motions is $H\Phi\left(\mathbf{R},\left\{ \mathbf{r}\right\} \right)=E\Phi\left(\mathbf{R},\left\{ \mathbf{r}\right\} \right)$
with Hamiltonian
\begin{eqnarray}
H & = & -\frac{\hbar^{2}}{2M}\nabla_{\mathbf{R}}^{2}+H\left(\left\{ \mathbf{r}\right\} \right)+V\left(\mathbf{R},\mathbf{\left\{ r\right\} }\right),
\end{eqnarray}
where the first term is the kinetic energy of the center-of-mass motion,
the second term describes the energy of the multiple relative motions
$\left\{ \mathbf{r}\right\} =\mathbf{r}_{1},\mathbf{r}_{2},\ldots$
and the third term is coupling between the center-of-mass motion and
the relative motions. For arbitrary center-of-mass space coordinate
$\mathbf{R}$, the eigenvalue equation
\begin{equation}
\left[H\left(\left\{ \mathbf{r}\right\} \right)+V\left(\mathbf{R},\mathbf{\left\{ r\right\} }\right)\right]\Theta_{k}\left(\mathbf{R},\mathbf{\left\{ r\right\} }\right)=E_{k}\left(\mathbf{R}\right)\Theta_{k}\left(\mathbf{R},\mathbf{\mathbf{\left\{ r\right\} }}\right)\label{eq:eigen}
\end{equation}
can be solved to obtain the corresponding eigenvalues $E_{k}\left(\mathbf{R}\right)$
and eigenfunctions $\Theta_{k}\left(\mathbf{R},\mathbf{\left\{ r\right\} }\right)$.
Since these bases $\left\{ \Theta_{k}\left(\mathbf{R},\mathbf{\mathbf{\left\{ r\right\} }}\right)\right\} $
are orthogonal and complete, one expands $\Phi\left(\mathbf{R},\left\{ \mathbf{r}\right\} \right)$
in bases $\left\{ \Theta_{k}\left(\mathbf{R},\mathbf{\mathbf{\left\{ r\right\} }}\right)\right\} $
as
\begin{eqnarray}
\Phi\left(\mathbf{R},\mathbf{\mathbf{\left\{ r\right\} }}\right) & = & \sum_{k=1}^{\infty}\Psi_{k}\left(\mathbf{R}\right)\Theta_{k}\left(\mathbf{R},\mathbf{\mathbf{\left\{ r\right\} }}\right).\label{eq:wave}
\end{eqnarray}
Obviously, this expanded wavefunction satisfies the original Schrördinger
equation as well. The straightforward derivation gives the set of
the effective motion equations of the coefficients $\Psi_{k}\left(\mathbf{R}\right)$
as
\begin{eqnarray}
H_{k}\left(\mathbf{R}\right)\Psi_{k}\left(\mathbf{R}\right)+\sum_{k'}H_{k,k'}^{1}\left(\mathbf{R}\right)\Psi_{k'}\left(\mathbf{R}\right) & = & E\Psi_{k}\left(\mathbf{R}\right),\label{eq:equation}
\end{eqnarray}
where
\begin{align}
H_{k}\left(\mathbf{R}\right) & =H_{k}^{0}\left(\mathbf{R}\right)+H_{k}^{1}\left(\mathbf{R}\right),\\
H_{k}^{0}\left(\mathbf{R}\right) & =-\frac{\hbar^{2}}{2M}\left(\nabla_{\mathrm{R}}-i\mathbf{A}_{k,k}\left(\mathbf{R}\right)\right)^{2}+E_{k}\left(\mathbf{R}\right),\\
H_{k}^{1}\left(\mathbf{R}\right) & =\sum_{k'\neq k}\frac{\hbar^{2}}{2M}\mathbf{A}_{k,k'}\left(\mathbf{R}\right)\cdot\mathbf{A}_{k',k}\left(\mathbf{R}\right),\\
H_{k,k'}^{1}\left(\mathbf{R}\right) & =i\frac{\hbar^{2}}{M}\sum_{k'\neq k}\mathbf{A}_{k,k'}\left(\mathbf{R}\right)\cdot\nabla_{\mathrm{R}}\Psi_{k'}\left(\mathbf{R}\right)\nonumber \\
 & +i\frac{\hbar^{2}}{2M}\sum_{k'\neq k}\int d\mathbf{r}\left[\nabla_{\mathrm{R}}\cdot\mathbf{A}_{k,k'}\left(\mathbf{R}\right)\right]\Psi_{k'}\left(\mathbf{R}\right)\\
 & +\frac{\hbar^{2}}{2M}\sum_{k'\neq k,k"}\mathbf{A}_{k,k"}\left(\mathbf{R}\right)\cdot\mathbf{A}_{k",k'}\left(\mathbf{R}\right)\Psi_{k'}\left(\mathbf{R}\right),\nonumber
\end{align}
and the Berry connections are defined as $\mathbf{A}_{k,q}\left(\mathbf{R}\right)\equiv i\int d\mathbf{r}\Theta_{k}^{*}\left(\mathbf{R},\mathbf{\mathbf{\left\{ r\right\} }}\right)\nabla_{\mathrm{R}}\Theta_{q}\left(\mathbf{R},\mathbf{\mathbf{\left\{ r\right\} }}\right).$
So far the effective motion equations are rigirous without any approximation.
Here, the $H_{k}^{0}\left(\mathbf{R}\right)$ and $H_{k}^{1}\left(\mathbf{R}\right)$
are adiabatic terms because they only involes the $k$-th energy-level.
However, $H_{k,k'}^{1}\left(\mathbf{R}\right)$ involve the transitions
between different energy-levles introducing the non-adiabatic processes.

To obtain the explicit expression for $\mathbf{A}_{k,q}\left(\mathbf{R}\right),$
differentiating the eigenvalue equation as Eq. (\ref{eq:eigen}) leads
to
\begin{align}
 & \left[H\left(\left\{ \mathbf{r}\right\} \right)+V\left(\mathbf{R},\mathbf{\left\{ r\right\} }\right)-E_{p}\left(\mathbf{R}\right)\right]\nabla_{\mathrm{R}}\Theta_{p}\left(\mathbf{R},\mathbf{\left\{ r\right\} }\right)\nonumber \\
 & =\left[\nabla_{\mathrm{R}}E_{p}\left(\mathbf{R}\right)-\nabla_{\mathrm{R}}V\left(\mathbf{R},\mathbf{\left\{ r\right\} }\right)\right]\Theta_{p}\left(\mathbf{R},\mathbf{\mathbf{\left\{ r\right\} }}\right).
\end{align}
Multiplying $\Theta_{k}^{*}\left(\mathbf{R},\mathbf{\left\{ r\right\} }\right)$
to both sides of the above equation and integrating over all relative
space coordinates $\mathbf{\left\{ r\right\} }$ gives

\begin{align}
 & \left[E_{k}\left(\mathbf{R}\right)-E_{p}\left(\mathbf{R}\right)\right]\int d\mathbf{r}\Theta_{k}^{*}\left(\mathbf{R},\mathbf{\mathbf{\left\{ r\right\} }}\right)\nabla_{\mathrm{R}}\Theta_{p}\left(\mathbf{R},\mathbf{\left\{ r\right\} }\right)\nonumber \\
 & =-\int d\mathbf{r}\Theta_{k}^{*}\left(\mathbf{R},\mathbf{\mathbf{\left\{ r\right\} }}\right)\nabla_{\mathrm{R}}V\left(\mathbf{R},\mathbf{\left\{ r\right\} }\right)\Theta_{p}\left(\mathbf{R},\mathbf{\mathbf{\left\{ r\right\} }}\right)
\end{align}
According to the definition of the Berry connection, the explicit
expression of the absolute value of the Berry connections is
\begin{align}
\left|\mathbf{A}_{k,p}\left(\mathbf{R}\right)\right| & =\left|\frac{\int d\mathbf{r}\Theta_{k}^{*}\left(\mathbf{R},\mathbf{\mathbf{\left\{ r\right\} }}\right)\left[\nabla_{\mathrm{\mathbf{R}}}V\left(\mathbf{R},\mathbf{\mathbf{\left\{ r\right\} }}\right)\right]\Theta_{p}\left(\mathbf{R},\mathbf{\mathbf{\left\{ r\right\} }}\right)}{E_{k}\left(\mathbf{R}\right)-E_{p}\left(\mathbf{R}\right)}\right|.\label{eq:42}
\end{align}
It is clear that the $H_{k}^{1}\left(\mathbf{R}\right)$ and $H_{k,k'}^{1}\left(\mathbf{R}\right)$
are regarded as the perturbations when the partial derivation of the
coupling $\nabla_{\mathrm{\mathbf{R}}}V\left(\mathbf{R},\mathbf{\mathbf{\left\{ r\right\} }}\right)$
is much smaller than the energy level spacing $\left|E_{k}\left(\mathbf{R}\right)-E_{p}\left(\mathbf{R}\right)\right|.$
The order of the perturbations can be characterized by the number
of the Berry connections. In this sense $H_{k}^{1}\left(\mathbf{R}\right)$
is the second order perturbation and $H_{k,k'}^{1}\left(\mathbf{R}\right)$
contains both the first order and the second order perturbations.

The Berry connections $\mathbf{A}_{k,k}\left(\mathbf{R}\right)$ in
the $H_{k}^{0}\left(\mathbf{R}\right)$ actually plays the role of
a gauge field. It is important to indicate that for Eq. (\ref{eq:eigen})
the phase of the bases $\left\{ \Theta_{k}\left(\mathbf{R},\mathbf{\mathbf{\left\{ r\right\} }}\right)\right\} $
are not fixed because the the eigenvalue equation is unchanged under
the transformation $\tilde{\Theta}_{k}\left(\mathbf{R},\mathbf{\mathbf{\left\{ r\right\} }}\right)=\Theta_{k}\left(\mathbf{R},\mathbf{\mathbf{\left\{ r\right\} }}\right)\exp\left[-i\theta\left(\mathbf{R}\right)\right]$.
However, the Berry connections of the transformed bases $\tilde{\mathbf{A}}_{k,q}\left(\mathbf{R}\right)\equiv i\int d\mathbf{r}\tilde{\Theta}_{k}^{*}\left(\mathbf{R},\mathbf{\mathbf{\left\{ r\right\} }}\right)\nabla_{\mathrm{R}}\tilde{\Theta}_{q}\left(\mathbf{R},\mathbf{\mathbf{\left\{ r\right\} }}\right)$
accordingly become
\begin{equation}
\tilde{\mathbf{A}}_{k,q}\left(\mathbf{R}\right)=\begin{cases}
\mathbf{A}_{k,q}\left(\mathbf{R}\right), & k\neq q\\
\mathbf{A}_{k,k}\left(\mathbf{R}\right)+\nabla_{\mathrm{R}}\theta\left(\mathbf{R}\right), & k=q
\end{cases}.
\end{equation}
Therefore $\mathbf{A}_{k,k}\left(\mathbf{R}\right)$ depends on the
choice of the phase factor $\theta\left(\mathbf{R}\right)$ and thus
we can not decide its perturbation order. This is actually the $U(1)$
gauge transformation and the physical observations are not influenced
by the specific choice of the phase factor. In our problem of the
interface exciton, this induced gauge field can be cancelled out by
choosing the proper bases as $\mathbf{A}_{k,k}\left(\mathbf{R}\right)=0$
for any $k$.

To apply the standard perturbation theory, we rewrite Eq. (\ref{eq:wave})
in a matrix form as
\begin{equation}
\Phi\left(\mathbf{R},\mathbf{\left\{ r\right\} }\right)=\Psi\left(\mathbf{R}\right)^{T}\cdot\Theta\left(\mathbf{R},\mathbf{\left\{ r\right\} }\right)\text{,}
\end{equation}
where the coefficient vector $\Psi\left(\mathbf{R}\right)$ and the
base vector $\Theta\left(\mathbf{R},\mathbf{r}\right)$ are
\begin{equation}
\Psi\left(\mathbf{R}\right)=\left[\begin{array}{c}
\Psi_{1}\left(\mathbf{R}\right)\\
\Psi_{2}\left(\mathbf{R}\right)\\
\vdots
\end{array}\right],\Theta\left(\mathbf{R},\mathbf{r}\right)=\left[\begin{array}{c}
\Theta_{1}\left(\mathbf{R},\mathbf{\left\{ r\right\} }\right)\\
\Theta_{2}\left(\mathbf{R},\mathbf{\left\{ r\right\} }\right)\\
\vdots
\end{array}\right].
\end{equation}
And the Eqs. (\ref{eq:equation}) are rewritten as $\left(H^{0}\left(\mathbf{R}\right)+H^{1}\left(\mathbf{R}\right)\right)\Psi\left(\mathbf{R}\right)=E\Psi\left(\mathbf{R}\right)$
with corresponding Hamiltonians in the matrix form as
\begin{equation}
H^{0}\left(\mathbf{R}\right)=\left[\begin{array}{ccc}
H_{1}^{0}\left(\mathbf{R}\right) & 0 & \cdots\\
0 & H_{2}^{0}\left(\mathbf{R}\right) & \cdots\\
\vdots & \vdots & \ddots
\end{array}\right]
\end{equation}
and
\begin{equation}
H^{1}\left(\mathbf{R}\right)=\left[\begin{array}{ccc}
H_{1}^{1}\left(\mathbf{R}\right) & H_{12}^{1}\left(\mathbf{R}\right) & \cdots\\
H_{21}^{1}\left(\mathbf{R}\right) & H_{2}^{1}\left(\mathbf{R}\right) & \cdots\\
\vdots & \vdots & \ddots
\end{array}\right].
\end{equation}
Here all the first order and the second order perturbations are included
into $H^{1}\left(\mathbf{R}\right)$. By applying the standard perturbation
theory, the second order eigen-energy and wavefunction respectively
as $E_{p}=E_{p}^{0}+E_{p}^{1}$ and $\Psi_{p}\left(\mathbf{R}\right)=\Psi_{p}^{0}\left(\mathbf{R}\right)+\Psi_{p}^{1}\left(\mathbf{R}\right),$
where
\begin{equation}
E_{p}^{1}=\sum_{k\neq p}\frac{\hbar^{2}}{2M}\int d\mathbf{R}\Psi_{p}^{0,*}\left(\mathbf{R}\right)\mathbf{A}_{p,k}\left(\mathbf{R}\right)\cdot\mathbf{A}_{k,p}\left(\mathbf{R}\right)\Psi_{p}^{0}\left(\mathbf{R}\right),\label{eq:48}
\end{equation}
\begin{align}
\Phi_{p}^{1}\left(\mathbf{R}\right) & =\sum_{k\neq p}\frac{\int d\mathbf{R'}\Psi_{k}^{\left(0\right),*}\left(\mathbf{R'}\right)H_{kp}^{1}\left(\mathbf{R}'\right)\Psi_{p}^{\left(0\right)}\left(\mathbf{R}'\right)}{E_{p}^{0}-E_{k}^{0}}\Psi_{k}^{0}\left(\mathbf{R}\right),
\end{align}
the zero-th order eigen-energy and wavefunction are determined by
$H^{0}\left(\mathbf{R}\right)$ as $H^{0}\left(\mathbf{R}\right)\Psi_{p}^{0}\left(\mathbf{R}\right)=E_{p}^{0}\Psi_{p}^{0}\left(\mathbf{R}\right)$
and $H_{kp}^{1}\left(\mathbf{R}'\right)$ is the element of perturbation
Hamiltonian $H^{1}\left(\mathbf{R}\right)$.

\begin{acknowledgements} The work is mainly supported by the Research Grant Council of Hong Kong (HKU705513P, C7036-17W), and the Croucher Foundation. Z. R. Gong is supported by NSFC Grants No. 11504241. \end{acknowledgements}

\end{document}